\documentclass[lettersize,journal]{IEEEtran}
\usepackage{amsmath,amsfonts}
\usepackage{algorithmic}
\usepackage{array}
\usepackage[caption=false,font=normalsize,labelfont=sf,textfont=sf]{subfig}
\usepackage{textcomp}
\usepackage{stfloats}
\usepackage{url}
\usepackage{verbatim}
\usepackage{graphicx}

\usepackage{booktabs} % For formal tables
\usepackage{stackengine}

\usepackage[table]{xcolor}
\usepackage{float}
\usepackage{xcolor}
\usepackage{wasysym}
\usepackage{colortbl}
\usepackage{soul}
\newcommand\encircle[1]{%
\tikz[baseline=(X.base)] 
  \node (X) [draw, scale=0.75, shape=circle, inner sep=0, fill=black, text=white, minimum size=0em] {\strut #1};}

\newcommand{\cmmnt}[1]{}  

\usepackage{caption}
\usepackage{mathtools}

\usepackage{tikz}

\usepackage{pifont}
\usepackage[noadjust]{cite}
\def\BibTeX{{\rm B\kern-.05em{\sc i\kern-.025em b}\kern-.08em
    T\kern-.1667em\lower.7ex\hbox{E}\kern-.125emX}}
\usepackage{balance}
\begin{document}
\title{PISA: A Binary-Weight Processing-In-Sensor Accelerator for Edge Image Processing}
\author{\IEEEauthorblockN{
Shaahin Angizi\IEEEauthorrefmark{2},
Sepehr Tabrizchi\IEEEauthorrefmark{1} and
Arman~Roohi\IEEEauthorrefmark{1}
}

\IEEEauthorblockA{\IEEEauthorrefmark{2}Department of Electrical and Computer Engineering, New Jersey Institute of Technology, Newark, NJ, USA\\
\IEEEauthorrefmark{1}School of Computing, University of Nebraska–Lincoln, Lincoln NE, USA\\
shaahin.angizi@njit.edu, aroohi@unl.edu} \vspace{-2em}}

% \author{Shaahin~Angizi,~\IEEEmembership{Member,~IEEE,}
%         Sepehr~Tabrizchi,~\IEEEmembership{Student Member,~IEEE,}
%         and Arman~Roohi,~\IEEEmembership{Member,~IEEE} % <-this % stops a space
%         \thanks{}
% \IEEEcompsocitemizethanks{\IEEEcompsocthanksitem S. Angizi is with Department of Electrical and Computer Engineering, New Jersey Institute of Technology, Newark, NJ, USA.
% E-mail: shaahin.angizi@njit.edu.
% \IEEEcompsocthanksitem S. Tabrizchi and A. Roohi are with School of Computing, University of Nebraska–Lincoln, Lincoln NE, USA.
% E-mail: aroohi@unl.edu.}
% }

\markboth{}
{How to Use the IEEEtran \LaTeX \ Templates}

\maketitle

\begin{abstract}
 This work proposes a Processing-In-Sensor Accelerator, namely PISA, as a flexible, energy-efficient, and high-performance solution for real-time and smart image processing in AI devices. PISA intrinsically implements a coarse-grained convolution operation in Binarized-Weight Neural Networks (BWNNs) leveraging a novel compute-pixel with non-volatile weight storage at the sensor side. This remarkably reduces the power consumption of data conversion and transmission to an off-chip processor. The design is completed with a bit-wise near-sensor processing-in-DRAM computing unit to process the remaining network layers. Once the object is detected, PISA switches to typical sensing mode to capture the image for a fine-grained convolution using only the near-sensor processing unit. Our circuit-to-application co-simulation results on a BWNN acceleration demonstrate acceptable accuracy on various image datasets in coarse-grained evaluation compared to baseline BWNN models, while PISA achieves a frame rate of 1000 and efficiency of $\sim$1.74 TOp/s/W. 
Lastly, PISA substantially reduces data conversion and transmission energy by $\sim$84\% compared to a baseline CPU-sensor design.
\end{abstract}


\section{Introduction}
\begin{figure}[t]
\begin{center}
\begin{tabular}{c}
\includegraphics [width=0.78\linewidth]{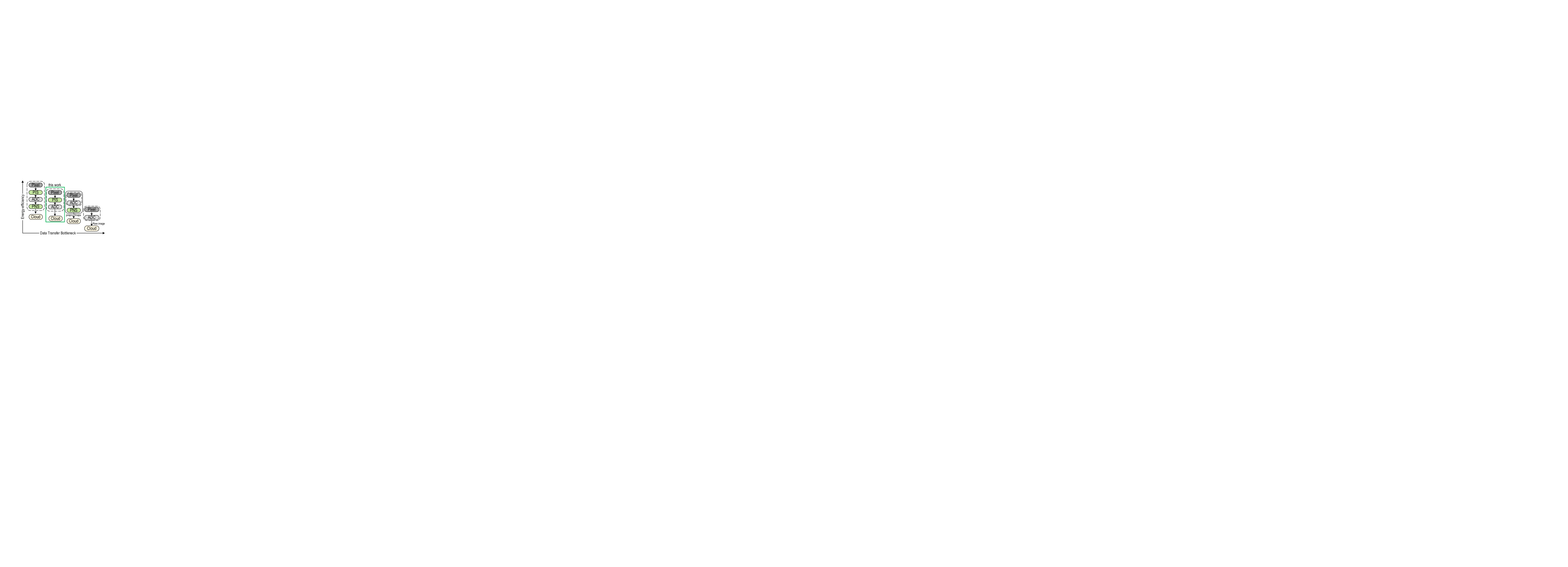}
\end{tabular}%\vspace{-0.75em}
\caption{Various visual system architectures.}
\label{intro}\vspace{-1.7em}
\end{center}
\end{figure}

\IEEEPARstart{I}{nternet} of Thing (IoT) devices are projected to attain an \$1100B market by 2025, with a web of interconnection projected to comprise approximately 75+ billion IoT devices, including wearable devices, smart cities, and smart industry \cite{hsu2019ai,yamazaki20174}. 
Intelligent IoT (IIoT) nodes consist of sensory systems, which enable massive data collection from the environment and people to process with on-/off-chip processors ($10^{18}$ bytes/s or ops). In most cases, large portions of the captured sensory data are redundant and unstructured. Data conversion and transmission of large raw data to a back-end processor impose high energy consumption, high latency, a memory bottleneck, and low-speed feature extraction on the edge \cite{hsu2019ai,gottardi201864,ko2017single} as shown with the pixel-only architecture in Fig. \ref{intro}. 
To overcome these issues, computing architectures will need to shift from a cloud-centric approach to a thing-centric (data-centric) approach, where the IoT node processes the sensed data.
Nonetheless, the processing demands of artificial intelligence tasks such as Convolutional Neural Networks (CNNs) spanning hundreds of layers face serious challenges for their tractability in computational and storage resources. Effective techniques in both software and hardware domains have been developed to improve CNN efficiency by alleviating the ``power and memory wall'' bottleneck.

In algorithm-based approaches, the use of shallower but wider CNN models, quantizing parameters, and network binarization has been explored thoroughly \cite{zhou2016dorefa,haj2018imaging,rastegari2016xnor}. Recently, low bit-width weights and activations reduces computing complexity and model size. For instance, in \cite{zhou2016dorefa}, authors performed bit-wise convolution between the inputs and low bit-width weights by converting the conventional Multiplication-And-Accumulate (MAC) into the corresponding AND-bitcount operations.
In an extreme quantization method, Binary Convolutional Neural Network (BCNN) has achieved acceptable accuracy on both small \cite{courbariaux2016binarized} and large datasets \cite{rastegari2016xnor} by relaxing the demands for some high precision calculations. Instead, it binarizes weight and/or input feature map while processing the forward path, providing a promising solution to mitigate aforementioned bottlenecks in storage and computational components \cite{angizi2017imc}.

\cmmnt{%%%%%%%%%%%%%%%%%%%%%%%%%%%%%%
Smart image sensors with instant image preprocessing have been explored for object recognition applications \cite{carey2013100,yamazaki20174,xu2020macsen,hsu20200}. This paves the way for new sensor paradigms such as a Processing-Near-Sensor (PNS), in which digital outputs of a pixel are accelerated near the sensor leveraging an on-chip processor. }%%%%%%%%%%%%%%%%%%%%%%%%%%%%%

From the hardware point of view, the underlying operations should be realized using efficient mechanisms. 
However, the conventional processing elements are developed based on the von-Neumann computing model with separate memory and processing blocks connecting via buses, which imposes serious challenges, such as long memory access latency, limited memory bandwidth, energy-hungry data transfer, and high leakage power consumption restricting the edge device's efficiency and working hours \cite{song2017pipelayer,yamazaki20174}.
Besides, in the upper level, this causes several significant issues such as communication bandwidth and security. 
Therefore, as a potential remedy, smart image sensors with instant image preprocessing have been extensively explored for object recognition applications \cite{carey2013100,yamazaki20174,xu2020macsen,hsu20200}. This paves the way for new sensor paradigms such as a Processing-Near-Sensor (PNS), in which digital outputs of a pixel are accelerated near the sensor leveraging an on-chip processor. 
Another solution to alleviate the above-mentioned challenges is a Processing-in-Memory (PIM) architecture, which is extensively studied in \cite{liu2020ns,li2017drisa,song2017pipelayer,zhou2020self}.
By inspiring the PNS and PIM techniques, two promising alternatives are the Processing-in-Sensor (PIS) that works on pre-Analog-to-Digital Converters (ADC) data \cite{xu2020macsen,xu2021senputing} and a hybrid PIS-PNS platform \cite{hsu2019ai} to improve vision sensor functionality and eliminate redundant data output, as shown in Fig. \ref{intro}. 
However, the computational capabilities of these sensors have been limited to specific applications since enhancing throughput is followed by a growth in sensor temperature; and higher temperatures lead to noise that degrades sensing accuracy \cite{chu2014neuromorphic}.
This includes specific feature extraction applications less supporting MAC-based image classification \cite{hsu2019ai,carey2013100} to meet both resiliency and efficiency such as Haar-like image filtering \cite{bong201714}, sharpening, blurring \cite{hsu20200}, and local binary pattern \cite{zhong20182pj}.

In this paper, we propose a new \underline{\textbf{P}}rocessing-\underline{\textbf{I}}n-\underline{\textbf{S}}ensor \underline{\textbf{A}}ccelerator (PISA) as an energy-efficient PIS paradigm co-integrating always-on sensing and processing capabilities working with a near-sensor PIM unit that is categorized as a new hybrid design. The proposed design features a real-time programmable coarse-grained convolution to reduce the power consumption of data conversion from photo-currents to pixel values in the image processing task. Once the object is detected, PISA switches to a typical sensing mode to capture the image for fine-grained convolution using a PNS unit.
The contributions of this paper are as follows:
\begin{enumerate}
    \item We develop a PIS architecture based on a set of innovative microarchitectural and circuit-level schemes optimized to process the 1$^{st}$-layer of BWNNs with weights stored in non-volatile memory components that offers energy-efficiency and speed-up.
    \item  We complete the design with a bit-wise near-sensor PIM-enabled unit based on DRAM to process the remaining network layers It leverages the charge-sharing feature of the DRAM cell and elevates it to implement the operation based on a dual-row activation mechanism.
    \item We present a solid bottom-up evaluation framework and a PIM assessment simulator to analyze the performance of the whole system.
    \item We extensively assess PISA's performance and energy-efficiency co-integrated with the near-sensor PIM unit compared with recent sensory platforms.%\vspace{-0.5em}
\end{enumerate}

The remainder of the paper is designed as follows. Section II discusses the state-of-the-art near-sensor and in-sensor processing designs and Magnetic Random Access Memory (MRAM). Section III delineates the proposed PISA architecture, and the supported operations and presents the near-sensor processing-in-DRAM solution. Section IV gives the proposed bottom-up evaluation framework and simulation results. Section V discusses the future work and finally, Section VI concludes this work. 

% To overcome the memory bandwidth bottleneck and address the existing challenges, we propose a  high-throughput and energy-efficient PIM accelerator based on DRAM, called \textit{DRIM}. \textit{DRIM} exploits a new in-memory computing mechanism called Dual-Row Activation (DRA) to perform bulk bit-wise operations between operands stored 

% \section{Binary-Weight Networks}
% The following equation \cite{courbariaux2015binaryconnect} shows the deterministic and stochastic binarization functions of floating-point weights $w_{fp}$: \vspace{-0.5em}

% \begin{equation}
% \footnotesize
% w_{b,De}= \left\{
% \begin{matrix}
% +1, w_{fp}\geq 0\\-1, w_{fp}<0
% \end{matrix}
% \right,
% w_{b,St}= \left\{
% \begin{matrix}
% +1,  \quad p=\sigma(w_{fp})\\
% \hspace{-2.6em}-1,\quad  1-p
% \end{matrix}
% \right.
% \end{equation}

% where $\sigma$ is a hard sigmoid function to determine the probability distribution: \vspace{-0.5em}

% \begin{equation}
% \footnotesize
% \sigma(x)=clip(\frac{x+1}{2},0,1)=max(0,min(1,\frac{x+1}{2}))
% \end{equation}

% Considering $k_i$ as the activation, $sign(\sum_{}^{}w_{i}k_i)$ gives the activated MAC operation.

\section{Background \& Motivation}
\subsection{Near-Sensor \& In-Sensor Processing} 
Systematic integration of computing and sensor arrays has been widely studied to eliminate off-chip data transmission and reduce ADC bandwidth by combining CMOS image sensor and processors in one chip as known as PNS \cite{li2021ns,hsu20200,bhowmik2019event,yamazaki20174,bong2017low,bhowmik2019visual}, or even integrating pixels and computation unit so-called PIS \cite{xu2020macsen,park20147,xu2021senputing,li20215,xu2020utilizing,xu20214}.
In \cite{hsu20200}, photocurrents are transformed into pulse-width modulation signals and a dedicated analog processor is designed to execute feature extraction reducing ADC power consumption. In \cite{yamazaki20174}, 3D-stacked column-parallel ADCs and Processing Elements (PE) are implemented to run spatiotemporal image processing. In \cite{kim2020chip}, a CMOS image sensor with dual-mode delta-sigma ADCs is designed to process  1$^{st}$-conv, layer of Binarized-Weight Neural Networks (BWNN). RedEye \cite{redeye} executes the convolution operation using charge-sharing tunable capacitors. Although this design shows energy reduction compared to a CPU/GPU by sacrificing accuracy, to achieve high accuracy computation, the required energy per frame increases dramatically by 100$\times$. 
MACSEN \cite{xu2020macsen} as a PIS platform processes the 1$^{st}$-conv. layer of BWNNs with the correlated double sampling procedure achieving 1000fps speed in computation mode. However, it suffers from humongous area-overhead and power consumption mainly due to the SRAM-based PIS method.
In \cite{pulsedomain}, a pulse-domain algorithm uses fundamental building blocks, photodiode arrays, and an ADC to perform near-sensor image processing that reduces design complexity and enhances both cost and speed. Putting all together, there are three main bottlenecks in IoT imaging systems that this work explores and aims to solve: (1) The conversion and storage of pixel values consume most of the power ($>$96\% \cite{choi2015energy,xu2020macsen}) in conventional image sensors; (2) the computation imposes a large area-overhead and power consumption in more recent PNS/PIS units and requires extra memory for intermediate data storage; and (3) the system is hardwired so the functionality is limited to simple pre-processing tasks such as 1$^{st}$-layer BWNN computation and cannot go beyond that.

\subsection{Processing-in-DRAM Platforms}
The PIM in the context of main memory (DRAM- \cite{li2017drisa,seshadri2017ambit,dai2018graphh}) has drawn much attention in recent years mainly due to larger memory capacities and off-chip data transfer reduction as opposed to SRAM-based PIM. Such processing-in-DRAM platforms show significantly higher throughput leveraging multi-row activation methods to perform bulk bit-wise operations by modifying the DRAM cell and/or SA. For example, Ambit \cite{seshadri2017ambit} uses Triple-Row Activation (TRA) method to implement majority-based AND/OR logic, outperforming Intel Skylake-CPU, NVIDIA GeForce GPU, and even HMC \cite{HMC} by 44.9$\times$, 32.0$\times$, and 2.4$\times$, respectively. DRISA \cite{driskill2011latest} employs 3T1C- and 1T1C-based computing mechanisms and achieves 7.7$\times$ speedup and 15$\times$ better energy-efficiency over GPUs to accelerate CNN.
However, there are various challenges in such platforms that make them inefficient acceleration solutions. (1) Given R=A$op$B  function ($op$ $\in$ {\tt AND2/OR2}), TRA-based method takes 4 consecutive steps to calculate one result as it relies on row initialization. Therefore TRA method needs an averagely 360$ns$ to perform such in-memory operations. Obviously, this row-initialization load could adversely impact the PIM's energy-efficiency; (2) By simultaneously activating three DRAM cells in TRA method or five cells in \cite{ali2019memory,angizi2019graphide}, the deviation on the Bit-Line is smaller than typical one-cell read operation in DRAM. This can elongate the sense amplification state or even adversely affect the reliability of the result.

\subsection{MRAM as a High-Performance Non-Volatile Memory}
With the great advancement of fabrication technology and commercialization of MRAM (e.g.,
IBM \cite{gallagher2006development} and Everspin \cite{everspin,EMD4E001G}), it is becoming a next-generation universal Non-Volatile Memory (NVM) technology, with potential applications in both last-level cache and main memory \cite{huai2008spin,kang2017modeling}.
Particularly, recent current-induced Spin-Transfer Torque (STT) and Spin-Orbit Torque (SOT)-based MRAMs have greatly changed the state-of-the-art memory hierarchy due to their non-volatility, zero leakage power in un-accessed bit-cell \cite{fukami2016spin,zhao2011sub}, high integration density (2$\times$ more than SRAM), high speed (sub-nanosecond) \cite{rowlands2011deep}, excellent endurance ($\sim10^{15}$ cycles \cite{kan2016systematic}), and compatibility with the CMOS fabrication process (back end of the line) \cite{fukami2016spin}. 
A standard 1-transistor 1-resistor (1T1R) STT-MRAM bit-cell consists of an access transistor and a Magnetic Tunnel Junction (MTJ). A typical MTJ structure consists of two ferromagnetic layers with a tunnel barrier sandwiched between them \cite{kawahara2011challenges}. One of the layers is a pinned magnetic layer, while the other one is a free magnetic layer. Due to the tunneling magnetoresistance (TMR) effect \cite{kawahara2011challenges}, the resistance of MTJ is high (/low) when the magnetization of two ferromagnetic layers are in anti-parallel (/parallel). The free layer magnetization could be manipulated by applying a current induced STT \cite{stiles2002anatomy}. Therefore, it is time for researchers to start in earnest to explore the application of MRAM in new energy-efficient in-memory and in-sensor computing systems that leverage its unique properties.

\begin{figure}[t]
\begin{center}
\begin{tabular}{c}
\includegraphics [width=0.92\linewidth]{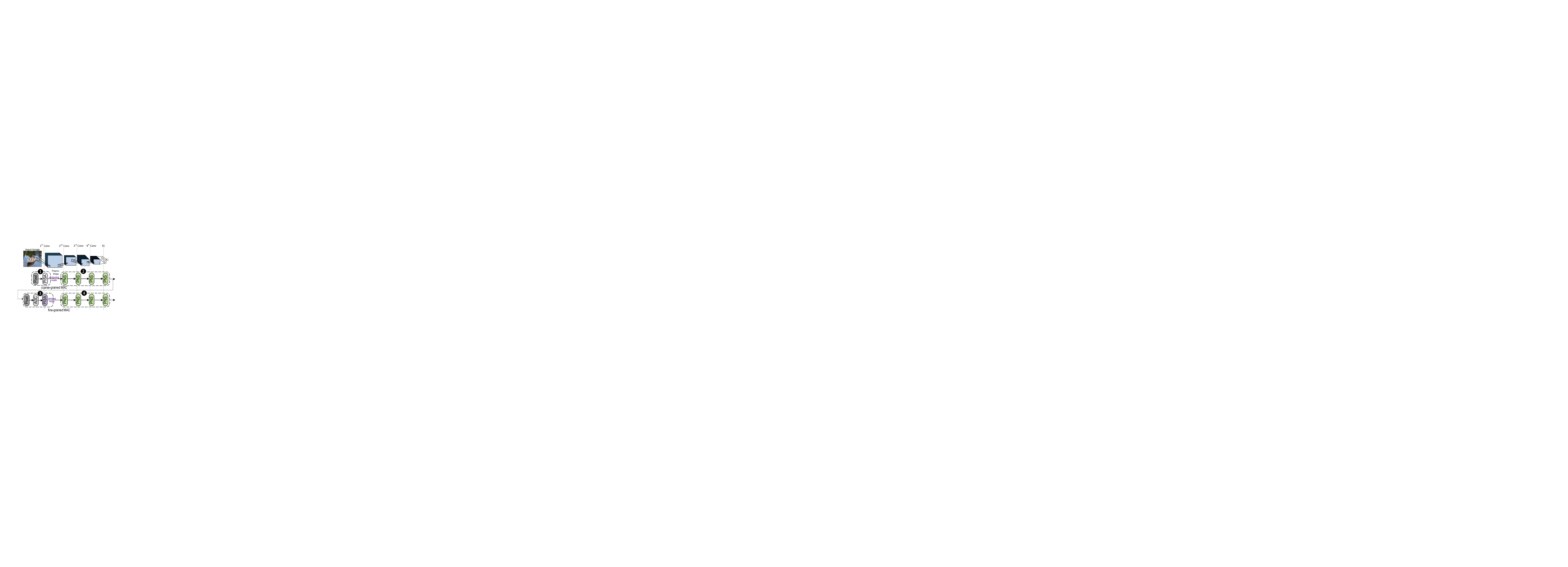}
\end{tabular}%\vspace{-0.75em}
\caption{Realizing coarse-grained and fine-grained computation in the proposed hybrid architecture.}
\label{proposed}\vspace{-2em}
\end{center}
\end{figure}

\section{Proposed Hybrid Processing-In-Sensor/ Near-Sensor Architecture}

Figure \ref{proposed} shows an overview of the proposed hybrid architecture's data flow regarding a simple network structure with four convolutional layers and one Fully-Connected (FC) layer. Similarly, our proposed approach can be extended to accelerate much more complex CNN models.
We first propose PISA as a flexible, energy-efficient, and high-performance solution for real-time and smart image processing in AI devices. PISA will integrate sensing and processing phases and can intrinsically implement a coarse-grained convolution operation (Fig \ref{proposed} \encircle{1}) required in a wide variety of image processing tasks such as \emph{classification} by processing the $1^{st}$-layer in BWNNs. The design will be completed with an on-chip reconfigurable PNS unit to perform a low bit-width coarse-grained convolution on the remaining layers. Once the object is roughly detected at the end of  step-\encircle{2}, PISA will switch to typical sensing mode \encircle{3} to capture the image for a fine-grained convolution using the near-sensor PIM unit \encircle{4}.

\begin{figure}[t]
\begin{center}
\begin{tabular}{c}
\includegraphics [width=0.95\linewidth]{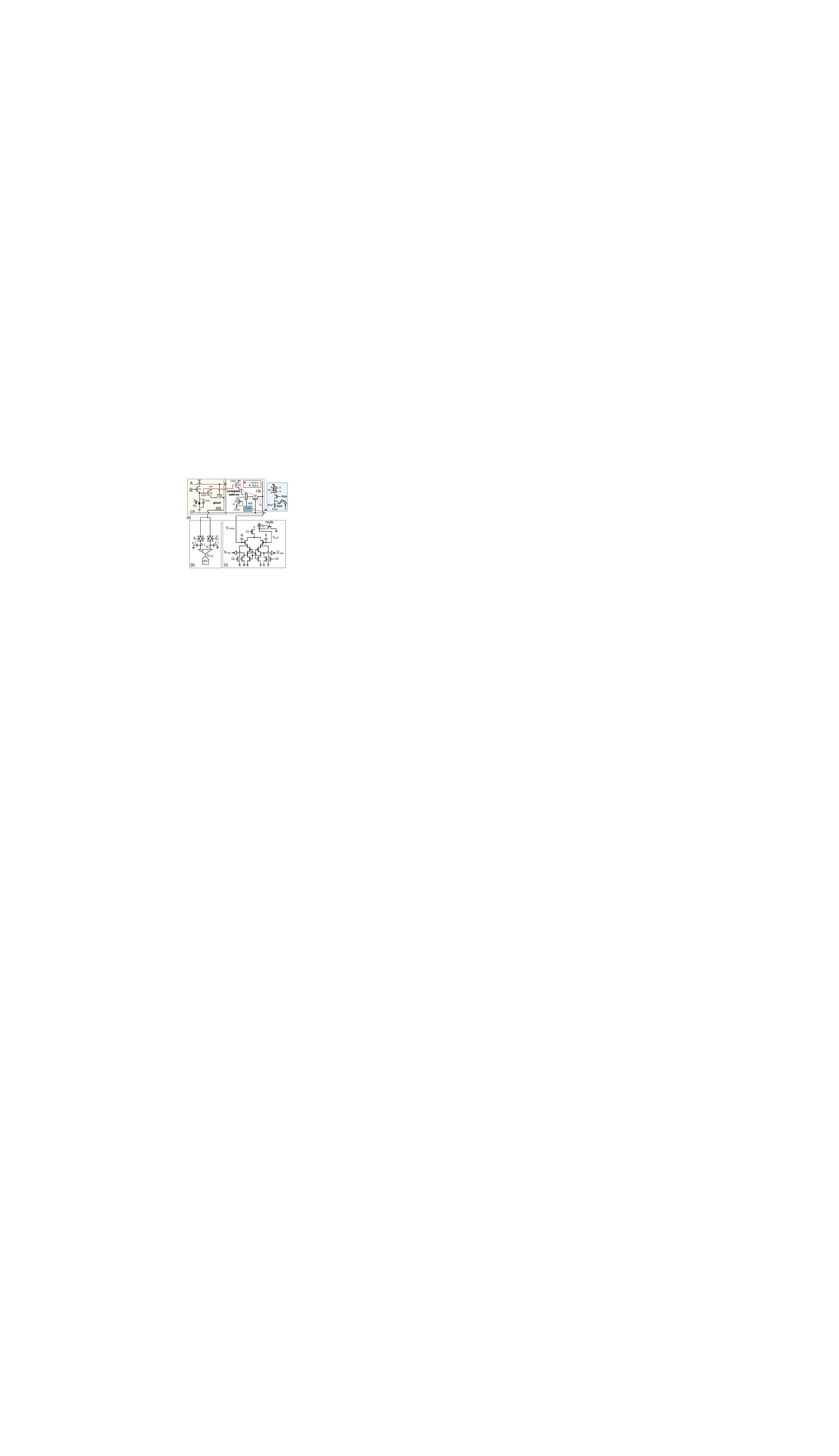}
\end{tabular}%\vspace{-1.5em}
\caption{(a) Compute-pixel element, (b) CP's read and conversion circuit in sensing mode realizing correlated double sampling procedure, (c) Sense amplifier design in processing mode based on StrongArm latch.}
\label{CP}%\vspace{-1.2em}
\end{center}
\end{figure}

\subsection{PISA Architecture}

\subsubsection{Compute-Pixel Element}
\begin{figure}[t]
\begin{center}
\begin{tabular}{c}
\includegraphics [width=0.96\linewidth]{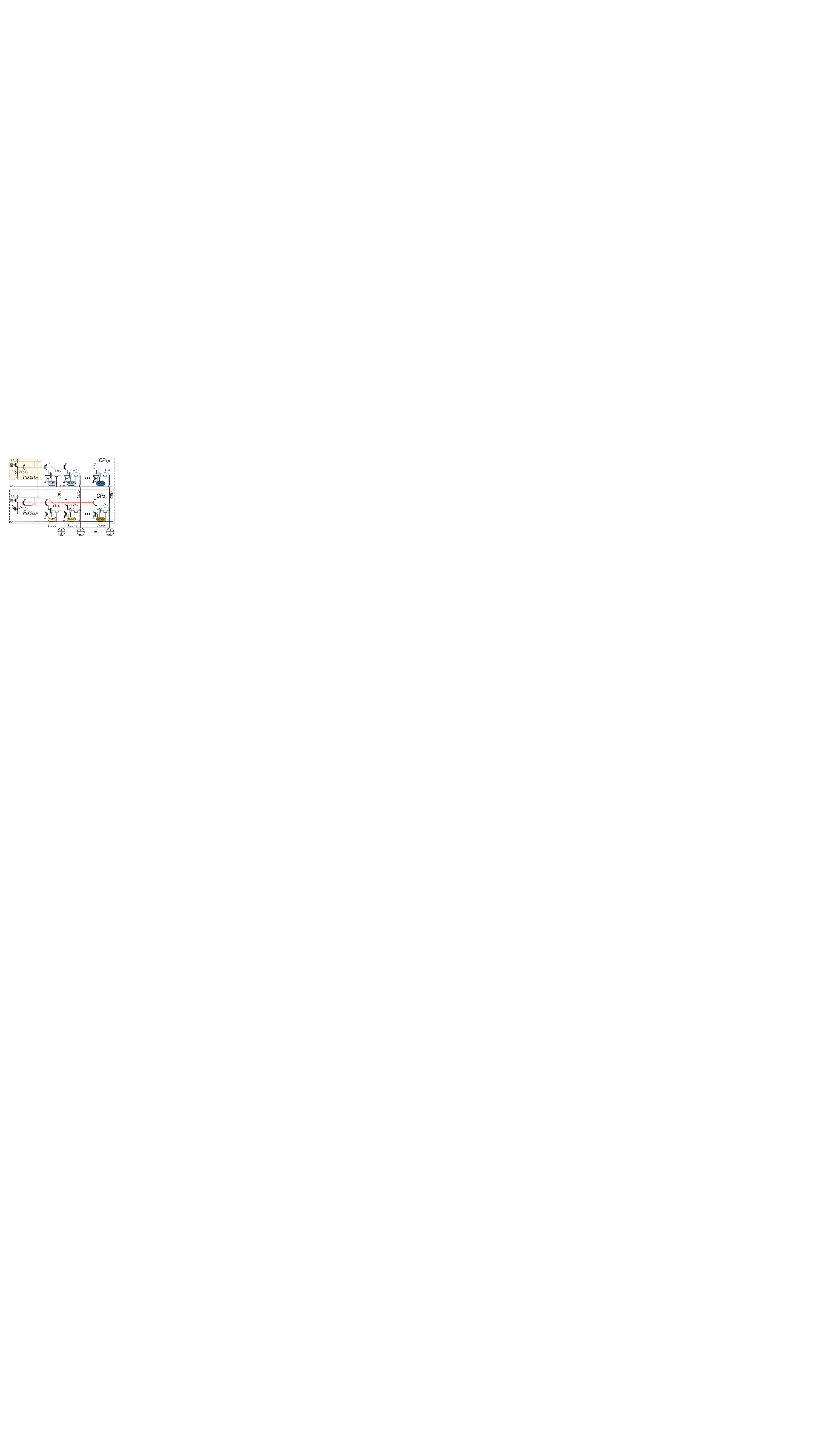}
\end{tabular}%\vspace{-1.5em}
\caption{A 2$\times$1 CP array in integrated sensing-processing mode.}
\label{CP2}%\vspace{-1.2em}
\end{center}
\end{figure}

To enable an integrated sensing and processing mode for PISA, we propose to upgrade the conventional pixel unit to a Compute-Pixel (CP). The CP is composed of a pixel (three transistors and one Photodiode (PD)) as shown in Fig. \ref{CP}(a), and $v$ compute add-ons. The compute add-on consists of three transistors of which T4 and T5 work as deep triode region current sources and a 2:1 MUX controlled by NVM element. We selected STT-MRAM as the NVM unit as depicted in Fig. \ref{CP}(a) due to its high speed (sub-nanosecond), long-endurance (10 years), and less than $fJ/bit$ memory write energy (close to SRAM) \cite{fong2011knack}. Thus, the binary weight data is stored as the magnetization direction in the MTJ's free layer, which could be programmed through the current-induced STT by NVM write driver. A reference resistor is then used to realize a voltage divider circuit to read out the weight value from the memory.  Fig. \ref{CP2} illustrates a 2$\times$1 CP array implementation. The \textit{Ri} (Row) signal is controlled by Row Ctrl and shared across CPs located in the same row to enable access during the row-wise sensing mode. However, the \textit{CR} (ComputeRow) is a unique controlling signal connected to entire CP units activated during processing mode. 
A Sense Bit-line (SBL) is shared across the pixels on the same column connected to sensor output for sensing-only mode (Fig. \ref{CP}(b)). Moreover, CPs share $v$ Compute Bit-lines (CBL), each connected to a SA for integrated sensing-processing mode.

\subsubsection{Operation Modes}
\begin{figure}[t]
\begin{center}
%\begin{tabular}{c}
\includegraphics [width=\linewidth]{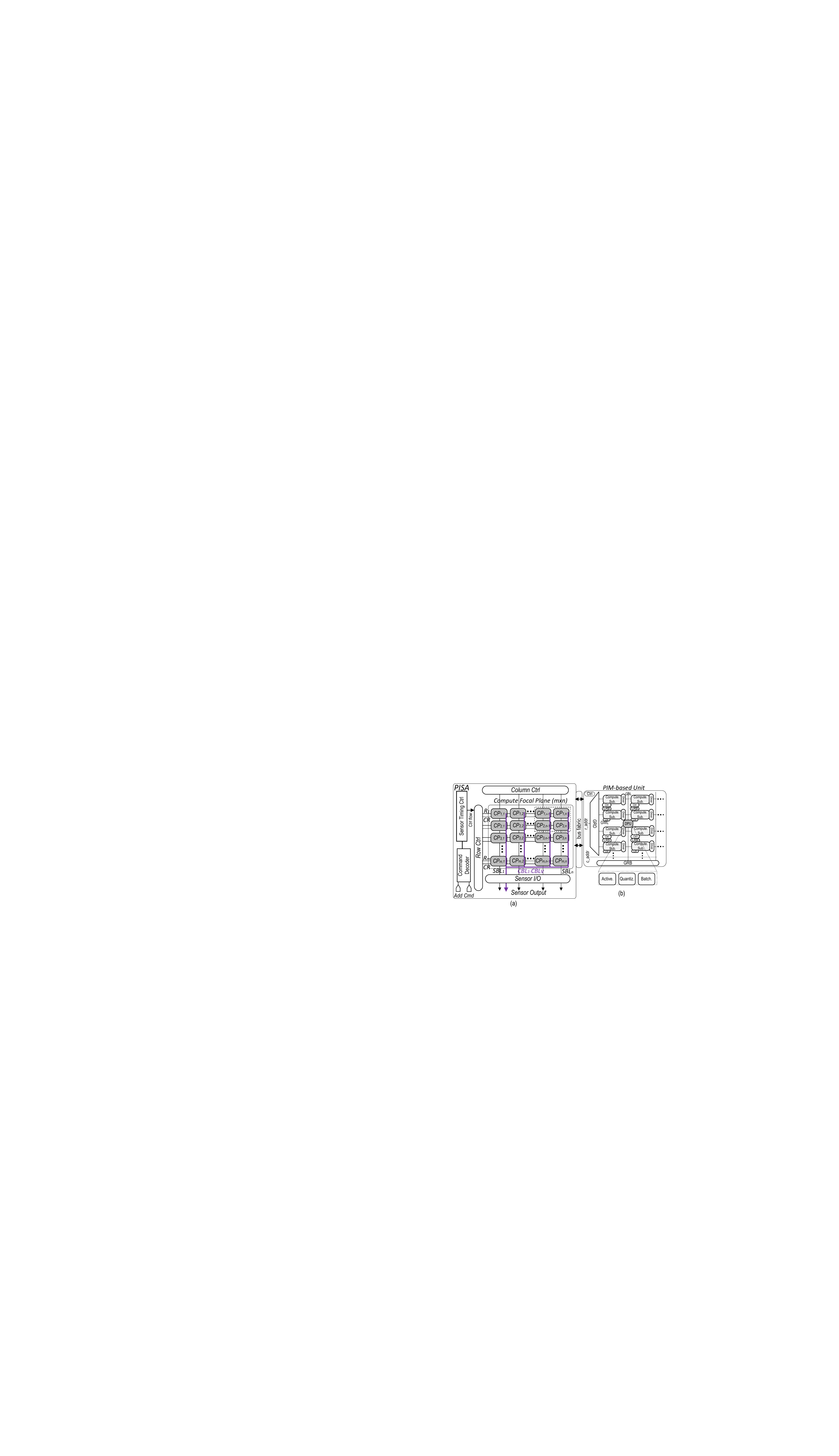}
%\end{tabular}%\vspace{-0.7em}
\caption{(a) The overview of PISA, and (b) PNS architectures.}
\label{arc}%\vspace{-1.7em}
\end{center}
\end{figure}

\begin{figure}[t]
\begin{center}
%\begin{tabular}{c}
\includegraphics [width=\linewidth]{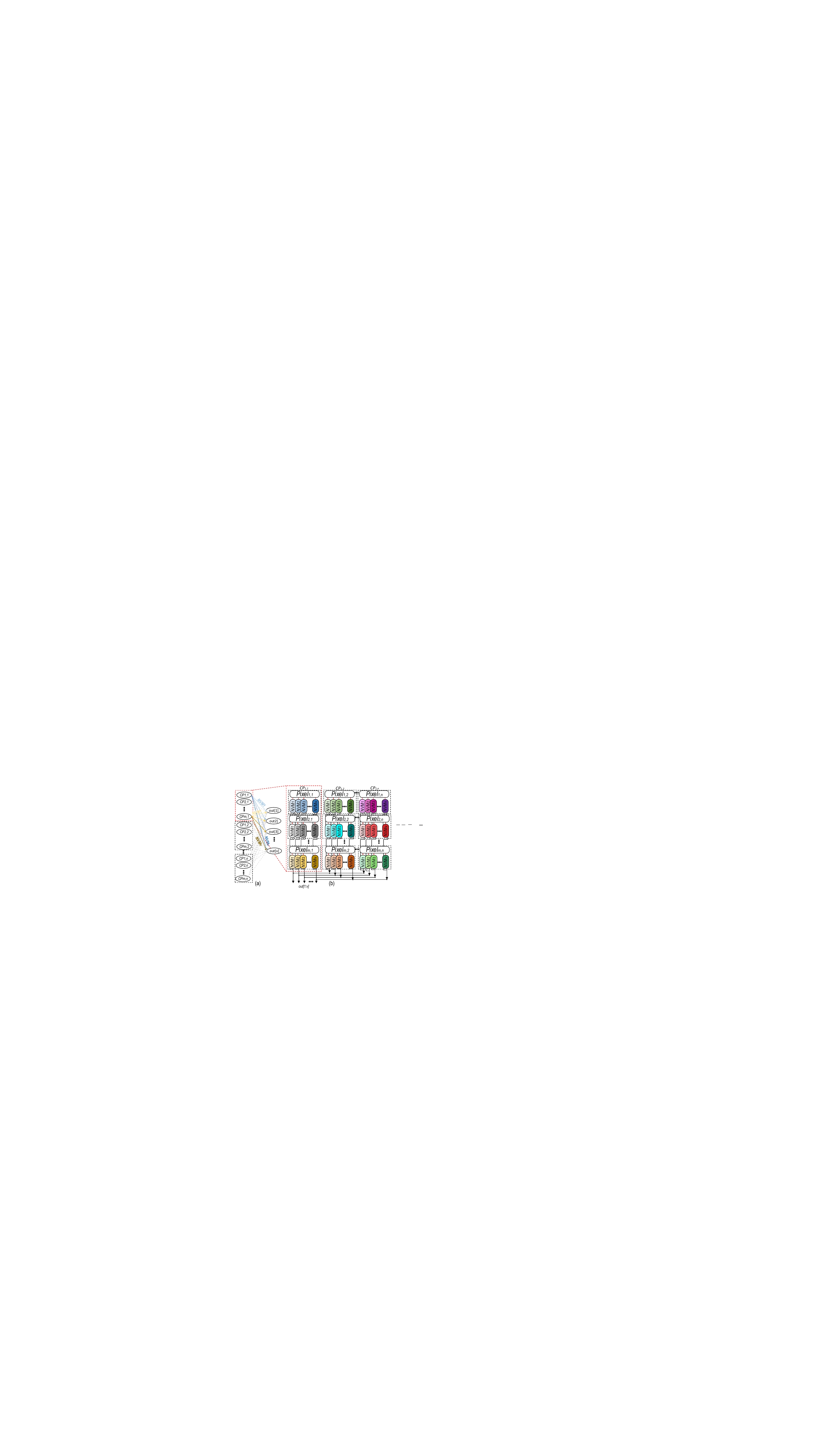}
%\end{tabular}
\vspace{0.3em}
\caption{(a) An example of fully-connected network with $v$ output, (b) PISA's mapping scheme for an m$\times$n CFP.}
\label{ex}\vspace{-1em}
\end{center}
\end{figure}

We develop PISA as a high-performance architecture for real-time and smart edge feature extraction as shown in Fig. \ref{arc}(a) on top of the proposed circuit schemes. At a high level, the PISA array consists of an $m\times n$ Compute Focal Plane (CFP), row and column controllers (Ctrl), command decoder, sensor timing Ctrl, and sensor I/O operating in two modes, i.e., sensing-only and integrated sensing-processing. The CFP is designed to co-integrate sensing and processing of the $1^{st}$-layer of BWNN targeting a low-power and coarse-grained detection. 
The $1^{st}$-layer binarized weight corresponding to each pixel is pre-stored into NVMs and an efficient coarse-grained MAC operation is then accomplished in a voltage-controlled cross-bar fashion (Fig. \ref{proposed} \encircle{1}). Accordingly, the output of the first layer is transmitted to a PNS or near-sensor PIM-based unit that enables the computation of the remaining BWNN layers \encircle{2}. Once the object is roughly detected at the edge, PISA switches to sensing mode like a traditional rolling-shutter CMOS image sensor \encircle{3}. % to transmit raw images to a near-sensor unit or off-chip processor for a fine-grained bit-wise convolution operation. 
It then transmits raw images to a near sensor unit \encircle{4}, or an off-chip processor, for a fine-grained bit-wise convolution operation. 
Fig. \ref{ex}(a) depicts a sample FC neural network, wherein CP$_{1,1}$-CP$_{m,n}$ are linked to out1 via NVM$_1$'s weight. Similarly, every pixel is connected to out2-out$v$. 
To maximize MAC computation throughput and fully leverage PISA's parallelism, we propose a hardware mapping scheme and connection configuration between CP elements and corresponding NVM add-ons shown in Fig. \ref{ex}(b) to implement the target neural network. In the following, the two operating modes of PISA are further elaborated.

\textit{Sensing Mode: }%\vspace{-0.25em}
In the sensing mode, by initially setting Rst=`high', the PD connected to the T1 transistor (see Fig. \ref{CP}(a)) turns into inverse polarization. In this way, turning on the access transistor T3  and k$_1$ switch (see Fig. \ref{CP}(b)) at the Sensor I/O allows the $C_1$ capacitor to fully charge through SBL. 
By turning off T1, PD generates a photo-current with respect to the external light intensity which in turn leads to a voltage drop ($V_{PD}$) at the gate of T2. Once again by turning on the T3 and this time k$_2$ switch, $C_{2}$ is selected to record the voltage drop. Therefore, the voltage values before and after the image light exposure, i.e., $V_{1}$ and $V_{2}$, are sampled by the CP, and the difference between two voltages is sensed with an amplifier. This value is proportional to the voltage drop on $V_{PD}$. In other words, the voltage at the cathode of PD can be read at the pixel output. Please note that in sensing mode, the CR signal is grounded. 
%\vspace{-1em}

\textit{Integrated Sensing-Processing Mode: }%\vspace{-0.25em}
In this mode, as shown in a sample 2$\times$1 CP array in Fig. \ref{CP2}, the $C_{PD}$ capacitor is initialized to the fully-charged state by setting Rst=`high', similar to the sensing mode.
During an evaluation cycle, by turning off T1, the row ctrl activates the CR signal, while the R$_i$ signals are deactivated. 
This will activate the entire array for a single-cycle MAC operation. The core idea behind compute add-on, shown in Fig. \ref{CP}(a), is to leverage pixel's $V_{PD}$ as a sampling voltage in $v$-NVM units to simultaneously generate (/pull) current through T4 (/T5) on the CBL. 
To implement multiplications between the pixel value identified by $V_{PD}$ and the binary weight stored in NVM, a 2:1 MUX unit was devised in every CP taking the T4 and T5 source signals as inputs and NVM sensed data as the selector. Note that T4 and T5 drains are connected to $V_{DD}$ and -$V_{DD}$, respectively. 
After exposure, the set of input sensor voltages $V_{PD}$= [$V_{{PD}_{1,1}}, V_{{PD}_{1,2}},..., V_{{PD}_{m,n}}$] is applied to the gate of T4s generating current set $I_{T4}$= [$I_{{1,1}_{(1)}}, I_{{1,1}_{(2)}},..., I_{{1,1}_{(v)}},..., I_{{m,n}_{(1)}}, I_{{m,n}_{(2)}},..., I_{{m,n}_{(v)}}$] for the entire array. If the binary weight equals `1' ($\textit{Wi}$=+1), T4 acts a current source and generates a current with  $I_{{i,j}_{(x)}}$ magnitude on the shared CBL as shown by the red dashed line in Fig. \ref{CP2}. However, if the binary weight equals `0' ($\textit{Wi}$=-1), T5 transistor acts a negative current source and pulls a current with the same magnitude as $I_{{i,j}_{(x)}}$ in the opposite direction from the shared CBL as indicated by the blue dashed line in Fig. \ref{CP2}.
This mechanism converts every input pixel value to a weighted current according to the NVM that is interpreted as the multiplication in BWNNs. Mathematically, let $G_{j,i}$ be the conductance of the synapse connecting $i^{th}$ to the $j^{th}$ node, the current through that synapse is $G_{j,i}V_i$ and the collection of the current through each CBL represents the MAC result ($I_{sum,j}$=$\sum_{i}^{}G_{j,i}V_i)$, according to Kirchhoff's law. This is readily calculated by measuring the voltage across a sensing resistor.
For the activation function, we designed and tuned a sense circuit connected to each CBL based on StrongARM latch to realize an in-sensor $sign$ function \cite{courbariaux2015binaryconnect,BNN1} as shown in Fig. \ref{CP}(c).
The SA requires two clock phases: pre-charge (Clk `high') and sensing (Clk `low').
During sensing, $I_{sum(x)}$ flows from every CBL to the ground and generates a sense voltage ($V_{sense}$) at the input of the SA. This voltage is compared with the reference voltage by applying a proportional current over a processing reference resistor ($R_{pro}$) activated by the mode signal. The binary activation is then transmitted through the bus fabrics to the PIM unit for storage.

\subsection{PNS Architecture}
Besides 1$^{st}$-layer, there are other convolutional and FC layers\footnote{FC can be equivalently implemented by convolution operations using $1\times1$ kernels \cite{zhou2016dorefa}.} in BWNNs that can be accelerated close to the sensor without sending the activated feature maps to off-chip processors.  
The general memory organization of the PNS is shown in Fig. \ref{arc}(b). The memory unit is divided into multiple banks consisting of computational sub-arrays. Every two sub-arrays share a Local Row Buffer (LRB) and the entire array shares a Digital Processing Unit (DPU) to pre-process the data by quantization and post-process outputs with linear batch normalization and activation. We divide the PNS's sub-array row space into two distinct regions as depicted in Fig. \ref{NEWSA}(a): 1- Data rows (500 rows out of 512) connected to a regular Row Decoder (RD), and 2- Computation rows (12), connected to a Modified Row Decoder (MRD), which enables two-row activation required for bulk bit-wise in-memory operations between operands. 

\subsubsection{Dual-Row Activation Mechanism}
\begin{figure}[t]
\begin{center}
\begin{tabular}{c}
\includegraphics [width=0.99\linewidth]{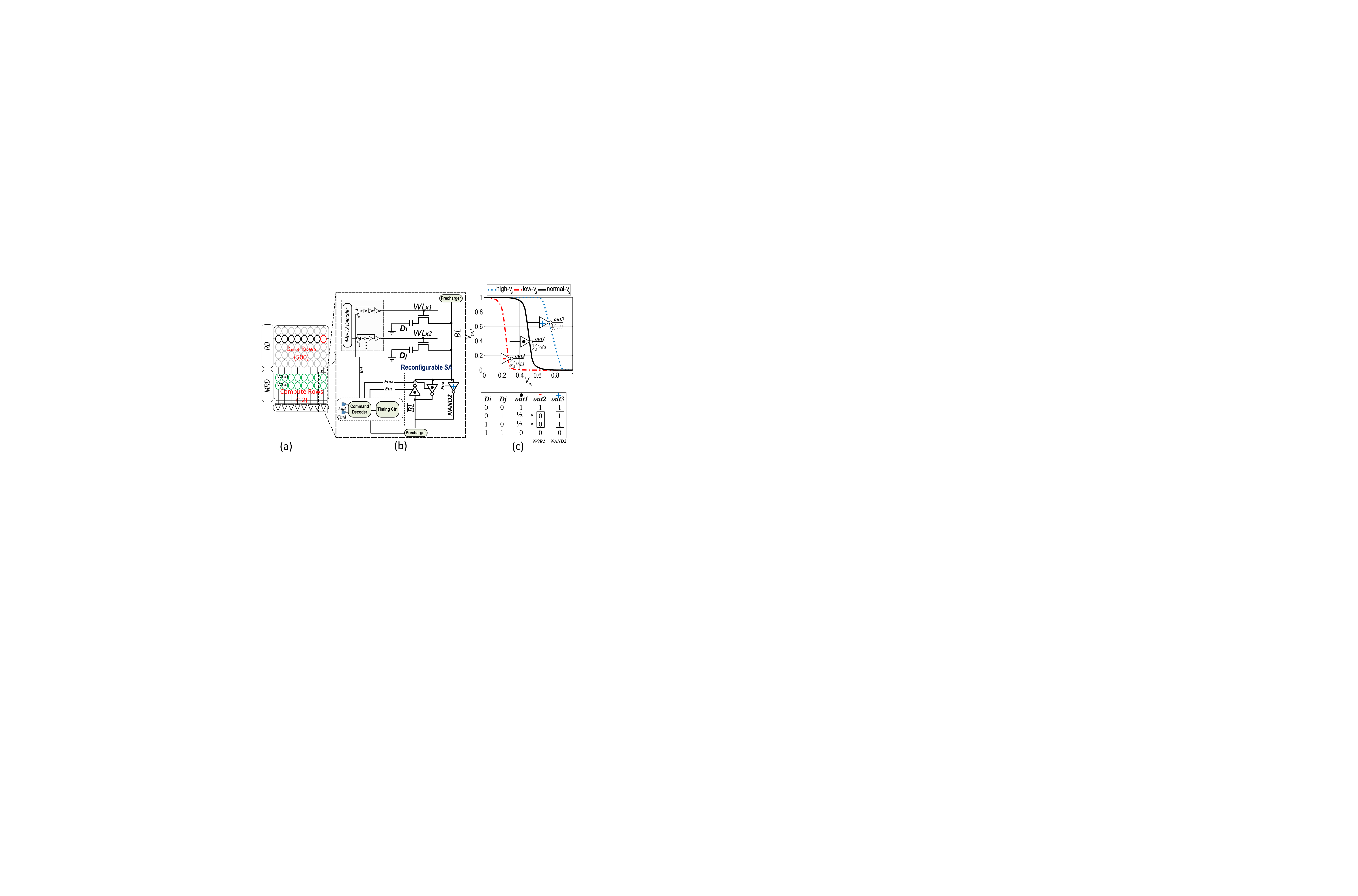}\\
 \end{tabular}\vspace{-0.8em}
\caption{(a) Sub-array organization, (b) DRA and the new sense amplifier design, (c) VTC and truth table of the SA's inverters.}
\label{NEWSA}
\end{center}\vspace{-0.2em}
\end{figure}

With careful observation of the existing processing-in-DRAM platforms, we realized that they impose reliability concerns and an excessive latency and energy to the memory chip, which could be alleviated by rethinking about SA circuit. Our key idea is to perform an in-memory {\tt NAND2} operation as a universal function through a Dual-Row Activation mechanism (DRA) to address these challenges. To achieve this goal, we propose a computational sub-array with new reconfigurable SA, as shown in Fig. \ref{NEWSA}(a)-(b), developed on top of the existing DRAM circuitry. The new SA consists of a regular DRAM SA with only one add-on inverter with three enable signals ($En_M$,$En_L$,$En_A$). This design leverages the charge-sharing feature of the DRAM cell and elevates it to implement {\tt (N)AND2} logic between two selected rows through static capacitive-NAND function in a single cycle. To implement capacitor-based logic, we use an inverter with shifted Voltage Transfer Characteristic (VTC), as shown in Fig. \ref{NEWSA}(c). In this way, a NAND logic can be readily carried out based on high switching voltage ($V_s$) inverter with standard high-$V_{th}$ NMOS and low-$V_{th}$ PMOS transistors. It is worth mentioning that, utilizing low/high-threshold voltage transistors along with normal-threshold transistors has been accomplished in the low-power application, and many circuits have enjoyed this technique in low-power design \cite{allam2000high,kuroda19960}.

\begin{figure}[t]
\begin{center}
\begin{tabular}{c}
\includegraphics [width=0.98\linewidth]{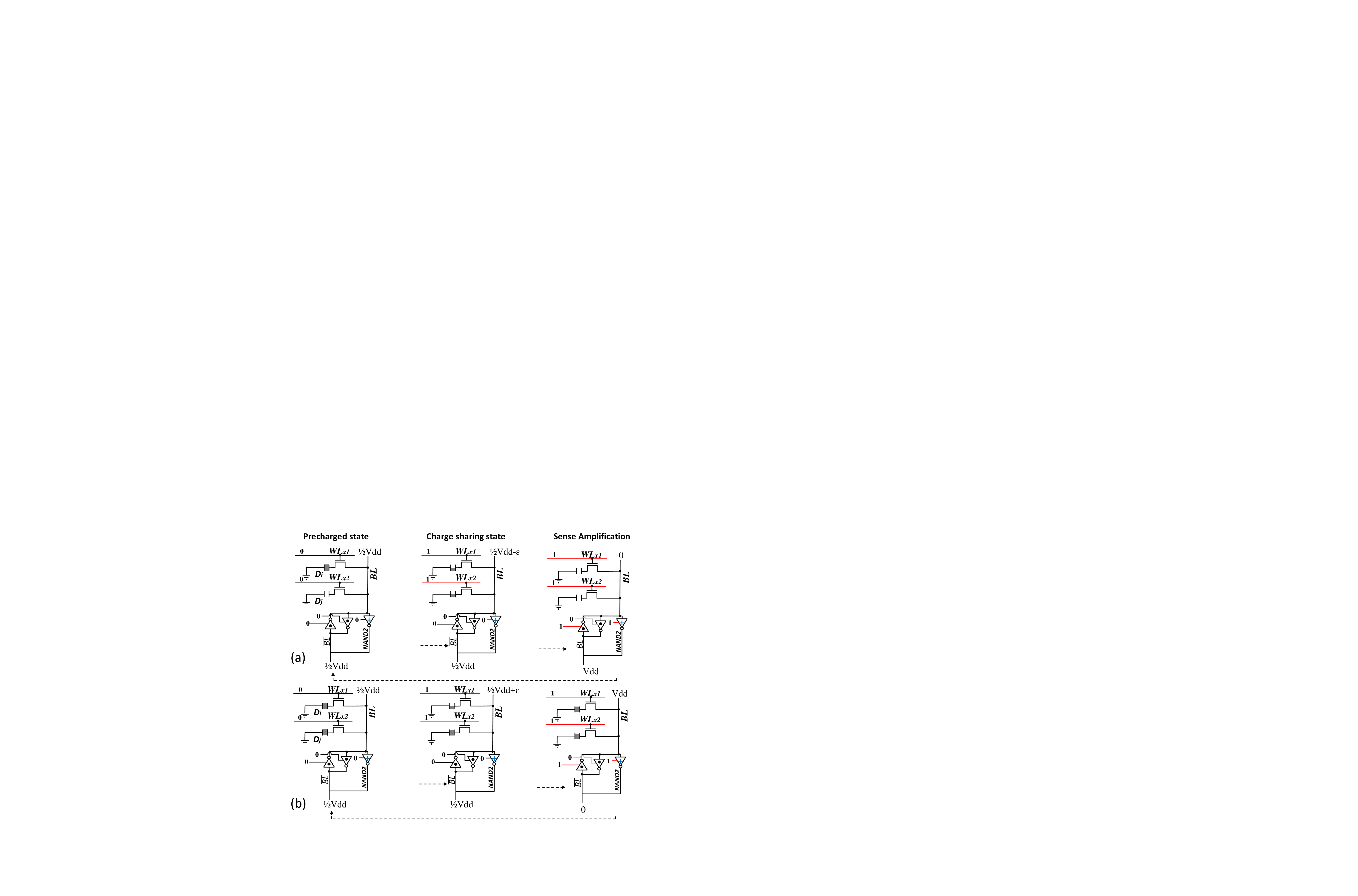}\\
 \end{tabular}\vspace{-0.8em}
\caption{Dual row activation to realize {\tt (N)AND2} for (a) ``10" and (b) ``11" input combinations.}
\label{AND}
\end{center}\vspace{-1.2em}
\end{figure}

To avoid original data overwritten as a common issue in processing-in-DRAM platforms \cite{seshadri2017ambit,angizi2019redram}, every operand row requires to be initially copied into compute rows before computation. Here, consider $D_i$ and $D_j$ operands are copied from data rows to $x1$ and $x2$ rows and both BL and $\overline{BL}$ are precharged to $\frac{V_{dd}}{2}$ (Precharged State in Fig. \ref{AND}).
To implement DRA, the Ctrl first activates two $WL$s in computational row space (here, $x1$ and $x2$) through the modified decoder for charge-sharing when all the other enable signals are deactivated (Charge Sharing State). 
During Sense Amplification State, by activating the corresponding enable signals ($En_L$ and $En_A$), the input voltage of high-$V_s$ inverter in the reconfigurable SA can be simply derived as  $V_i=\frac{n.V_{dd}}{C}$, where $n$ is the number of DRAM cells storing logic `1' and $C$ represents the total number of unit capacitors connected to the inverter (i.e., 2 in DRA method). Now, the high-$V_{s}$ inverter amplifies the deviation from $\frac{3}{4}V_{dd}$ and realizes a {\tt NAND2} function and writes back the inverted result in a single memory cycle.

\subsubsection{Hardware Mapping}
\begin{figure}[b]
\begin{center}
%\begin{tabular}{l}
\includegraphics [width=\linewidth,height=6.5cm]{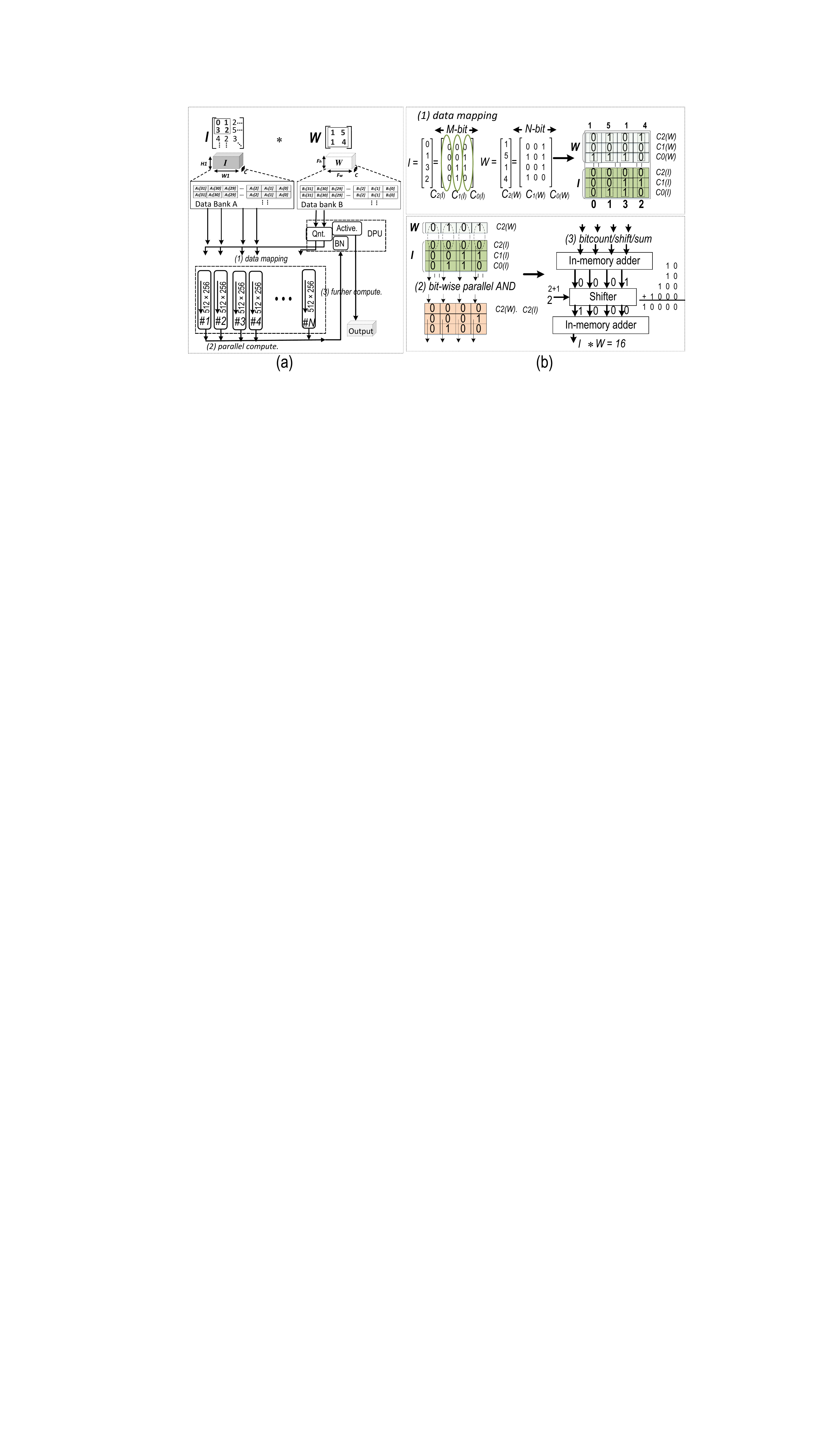}% \vspace{-0.8em}\\
% \end{tabular}%\vspace{-0.4em}
\caption{Acceleration steps of PNS convolver.}
\label{PIM}\vspace{-2em}
\end{center}
\end{figure}

Figure \ref{PIM} gives an overview of the proposed BWNN bit-wise acceleration steps.
In the first step, the preprocessed data from PISA is mapped into the PNS's computational sub-arrays. In the second step, parallel computational sub-arrays, which are designed to handle the computational load employing PIM techniques, perform bulk bit-wise operations between tensors and generate the output. Accordingly, the output is activated by DPU's activ. unit and saved back into memory.
From a computation perspective, every conv. layer can be similarly implemented by exploiting \textit{logic AND}, \textit{bitcount}, and \textit{bitshift} as rapid and parallelizable operations \cite{zhou2016dorefa}.
Assume $I$ is a sequence of $M$-bit input integers (3-bit as an example in Fig. \ref{PIM}) located in input fmap covered by sliding kernel of $W$, such that $I_i \in I$ is an $M$-bit vector representing a fixed-point integer. 
Now, we index the bits of each $I_i$ element from LSB to MSB with $m=[0 , M-1]$, such that $m=0$ and $m=M-1$ are corresponding to LSB and MSB, respectively. Accordingly, we represent a second sequence denoted as $C_m(I)$ including the combination of $m^{th}$ bit of all $I_i$ elements (shown by colored elliptic). For instance, $C_0(I)$ vector consists of LSBs of all $I_i$ elements ``0110". Considering $W$ as a sequence of $N$-bit weight integers (3-bit, herein) located in a sliding kernel with index of $n=[0 , N-1]$. The second sequence can be similarly generated as $C_n(W)$. Now, by considering the set of all $m^{th}$ value sequences, the $I$ can be represented like $I=\sum_{m=0}^{M-1}2^mc_m(I)$. Likewise, $W$ can be represented like $W=\sum_{n=0}^{N-1}2^nc_n(W)$. In this way, the convolution between $I$ and $W$ can be defined as $\small \sum _{m=0}^{M-1}\sum _{n=0}^{N-1}2^{m+n}bitcount(and(C_n(W),C_m(I)))$. 
As shown in the data mapping step of Fig. \ref{PIM}, $C_2(W)$-$C_0(W)$ are consequently mapped to the designated sub-array. Accordingly, $C_2(I)-C_0(I)$ are mapped in the following memory rows in the same way. Now, computational sub-array can perform bit-wise parallel AND operation of $C_n(W)$ and $C_m(I)$ as depicted in Fig. \ref{PIM} leveraging the DRA mechanism. The results stored within the sub-array will be accordingly processed using DPU's bit-counter. Bit-counter readily adds up the number of ``1''s in each resultant vector and passes it to the Shifter unit. As depicted in Fig. \ref{PIM}, ``0001", as result of Bit-Counter is left-shifted by 3-bit ($\times2^{2+1}$) to ``1000". Eventually, the PIM adds the shifter unit's outputs to produce output fmaps for every layer. Note that the PNS supports multi-bit convolution so the various configurations of weight:input can be achieved at the edge.

\section{Performance Evaluation}

\subsection{Framework \& Methodology}
\begin{figure}[t]
\begin{center}
\begin{tabular}{l}
\includegraphics [width=0.99\linewidth]{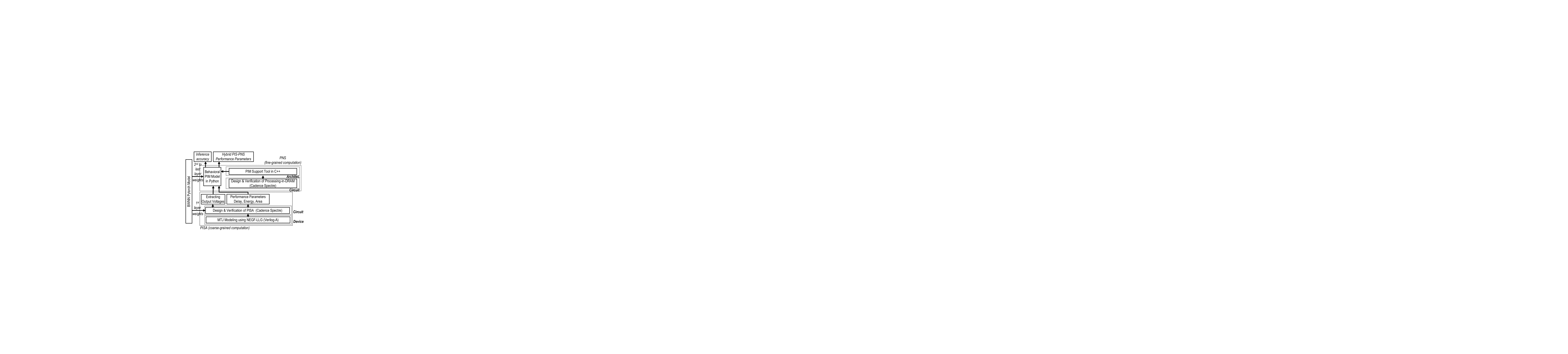} %\vspace{-0.1em}\\
 \end{tabular}%\vspace{-0.4em}
\caption{Evaluation framework.}
\label{Evalf}\vspace{-1em}
\end{center}
\end{figure}

To assess the performance of the proposed design, we developed a simulation framework from scratch consisting of two main components as shown in Fig. \ref{Evalf}.
\underline{First}, for coarse-grained computation, at the circuit level, we fully implemented PISA with peripheral circuity with TSMC 65nm-GP in Cadence to achieve the performance parameters. For the NVM elements, we jointly use the Non-Equilibrium Green's Function (NEGF) and Landau-Lifshitz-Gilbert (LLG) equations to model MTJ \cite{fong2011knack}. A Verilog-A model of NVM element is then developed to co-simulate with interface CMOS circuits in Cadence Spectre and SPICE.
PISA requires binarizing the 1$^{st}$-layer weights as discussed while the rest of the layers processed with the PIM unit have various bit-length. We trained a PyTorch BWNN model inspired by \cite{FBNA, tang2017train} extracting the 1$^{st}$-layer weights. PISA's NVM elements are then programmed at the circuit-level by the binary weights. After  1$^{st}$-layer computation, the results are recorded and fed into a behavioral-level PIM simulator to simulate the near-sensor PIM platform. 
\underline{Second}, for fine-grained computation, at the circuit level, we fully implemented the PNS and DRISA-1T1C \cite{li2017drisa} with TSMC 65nm-GP in Cadence to achieve the performance parameters. An architecture-level PIM support tool is developed to model the timing, energy, and area based on the circuit-level data. This tool offers the same flexibility in memory configuration regarding bank/mat/subarray organization and peripheral circuitry design as Cacti \cite{thoziyoor2008cacti} while supporting PIM-level configurations. Based on the circuit level results, we altered the configuration files (.cfg) with different array organizations and add-ons such as DPU and achieved performance for PIM operations. We then configured the PIM unit with 1024 rows and 256 columns, 4$\times$4 mats per bank organized in an H-tree routing manner, and 16$\times$16 banks (with 1/1 as row/column activation) in each memory group.
The behavioral PIM model developed in Python then takes coarse-grained computation voltage results, 2$^{nd}$-to-last layer trained weights, and the PIM architecture-level data and processes the BWNN. It calculates the latency and energy that the whole system spends executing the network.

\subsection{Functionality}
\begin{figure}[t]
\begin{center}
%\begin{tabular}{l}
\includegraphics [width=\linewidth]{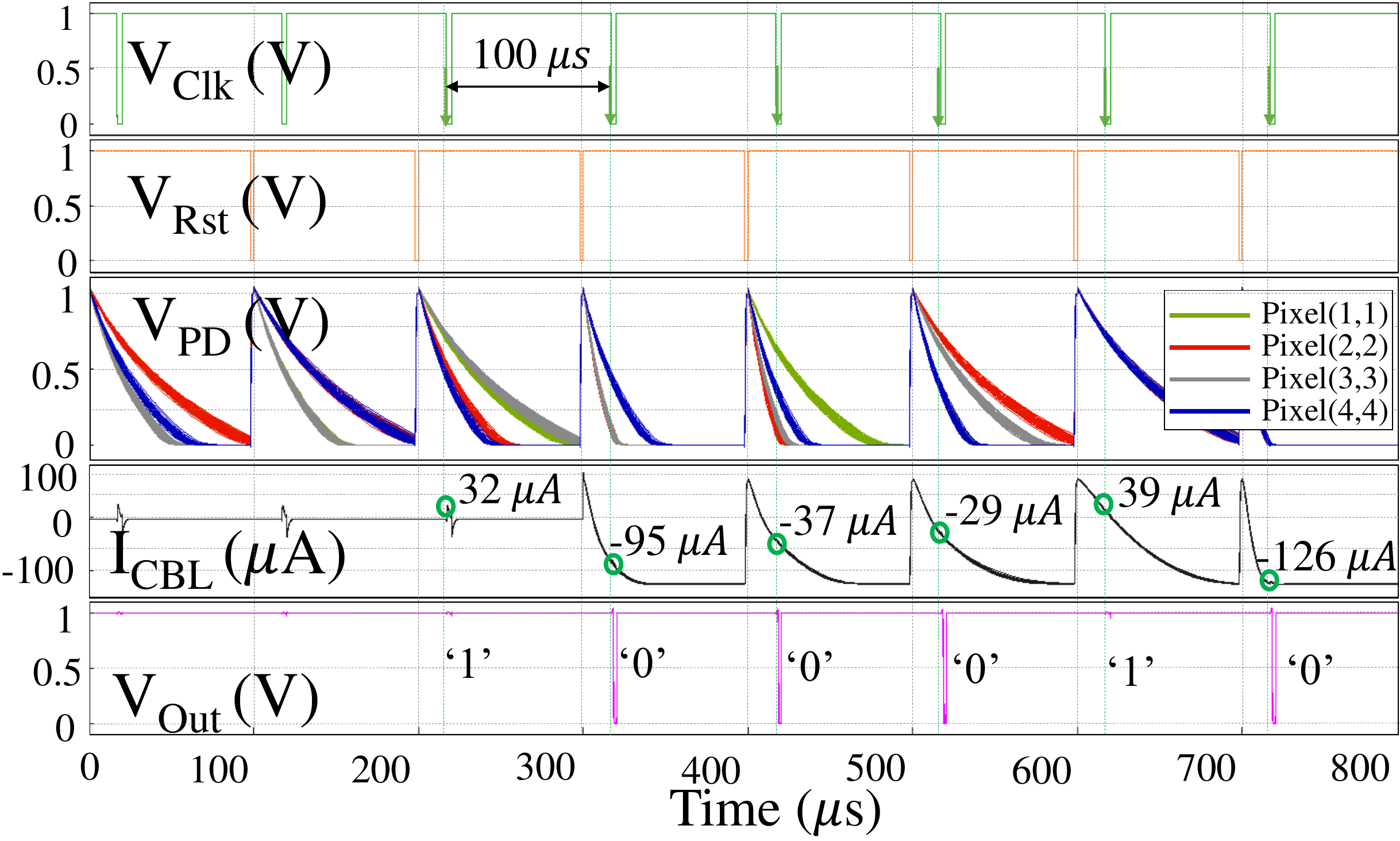} %\vspace{-0.8em}\\
% \end{tabular}%\vspace{-0em}
\caption{Post-layout transient simulation result for a sample 4$\times$4 PISA array.}
\label{wave}\vspace{-1em}
\end{center}
\end{figure}

Fig. \ref{wave} shows the post-layout transient simulation waveforms of a 4$\times$4 PISA array with eight NVM units ($v$=8) storing binary weights with $V_{Clk}$, $V_{Rst}$, $V_{PD}$, $I_{CBL}$, and $V_{Out}$ signals. PISA executes global shutter in processing mode and conducts all computations in parallel. As shown, periodically, by precharging $V_{PD}$ to VDD, the computation takes place at every falling edge of the clock, i.e., $\sim$100$\mu$s. In this way, $I_{CBL}$ carries the summation current corresponding to $V_{PD}$s. As can be seen, when $I_{CBL}$ is positive (e.g., the case of 32$\mu$A and 39$\mu$A) meaning the MAC result is larger than zero and the output $sign$ function results in ``1'' and vice-versa.

\begin{figure}[b]
\begin{center}
\begin{tabular}{c}
\includegraphics [width=0.98\linewidth]{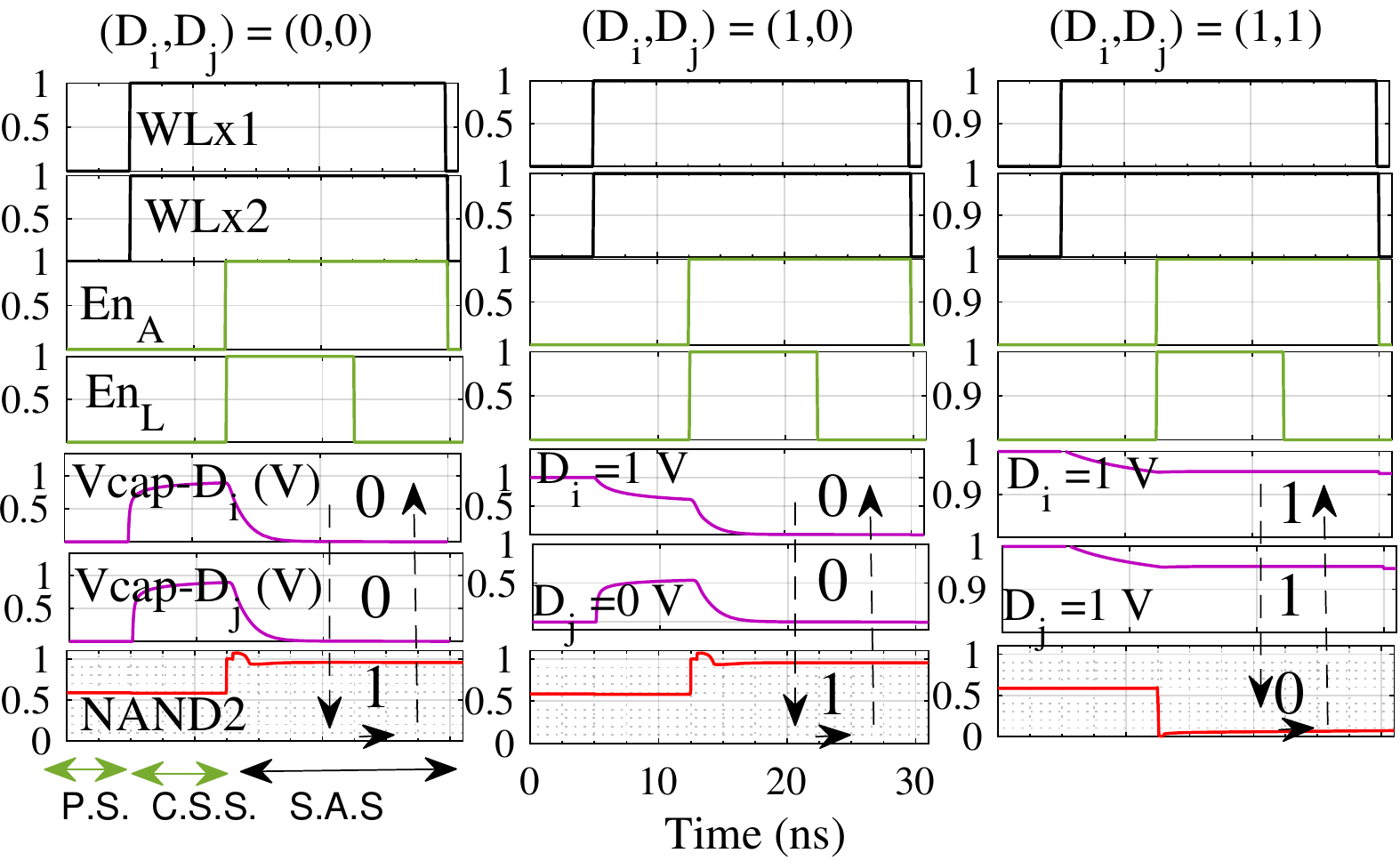}\\
 \end{tabular}\vspace{-0.8em}
\caption{The transient simulation of a single PNS sub-array. P.S., C.S.S., S.A.S. are short for Precharged State, Charge Sharing State, and Sense Amplification State, respectively.}
\label{tran}
\end{center}
\end{figure}

The transient simulation results of the in-DRAM DRA mechanism to realize single-cycle {\tt (N)AND2} operation is shown in Fig. \ref{tran} for three possible input combinations. We can observe how NAND output and accordingly cell's capacitor is charged to $V_{dd}$ (when $D_iD_j$=11) or discharged to GND (when $D_iD_j$=00/01/10) during sense amplification state.

\subsection{Robustness}
We simulated the PISA's circuit-level variations and noises with equivalent post-layout parasitic at 300K with 10000 Monte-Carlo runs. This includes a variation in width/length of transistors and CBL capacitance. Besides, the impact of thermal noises was modeled as the additive Gaussian noise on the dynamic capacitance along with 1/f noise of CMOS transistors from the source-follower in pixels. Our study shows that percentage of failure upon a considerable variation/noise (10\%) across 10000 iterations is 0\% as plotted $V_{PD}$ in Fig. \ref{wave}.
For variations above 10\%, a noise-aware training technique is used injecting multiplicative noise onto the weights in the training to increase BWNN robustness. For the NVM element, we added a $\sigma = 2\%$ variation to the Resistance-Area product, and a $\sigma = 5\%$ process variation (typical MTJ conductance variation \cite{fong2011knack}) on the TMR and verified a sense margin of 70mV between parallel and anti-parallel cases.

As for PNS unit, we performed a comprehensive circuit-level simulation to study the effect of process variation on both DRA and TRA methods considering different noise sources and variation in all components including DRAM cell (BL/WL capacitance and transistor, shown in Fig. \ref{cap}) and SA (width/length of transistors-$V_s$). We ran Monte-Carlo simulation (DRAM cell parameters were taken and scaled from Rambus \cite{Rambus}) under 10000 trials and increased the amount of variation from $\pm$0\% to $\pm$30\% for each method. Table \ref{var} shows the percentage of the test error in each variation. We observe that even considering a significant $\pm$10\%  variation, the percentage of erroneous DRA across 10000 trials is 0\%, where the TRA method shows a failure with 0.18\%. 

\begin{table}[h]
\begin{minipage}[b]{0.5\linewidth}
\centering
\includegraphics[width=0.8\linewidth]{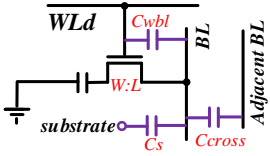}
\vspace{-0.5em}
    \captionof{figure}{Noise sources in DRAM cell. Glossary: $Cwbl$, $Cs$, and $Ccross$ are WL-BL, BL-substrate, and BL-BL capacitance, respectively.}
    \label{cap}
\end{minipage}
\begin{minipage}[b]{0.43\linewidth}
\centering
    \small\vspace{-1em}
\begin{tabular}{|c|c|c|}
\hline
Variation & TRA  & DRA  \\ \hline
$\pm$5\%  & 0.00 & 0.00 \\ \hline
$\pm$10\% & 0.18 & 0.00 \\ \hline
$\pm$15\% & 5.5  & 1.2  \\ \hline
$\pm$20\% & 17.1 & 9.6  \\ \hline
$\pm$30\% & 28.4 & 16.4  \\ \hline
\end{tabular}\vspace{3em}
    \caption{Process variation analysis.}\vspace{-1em}
    \label{var}
\end{minipage}\hfill
\end{table}

\subsection{Energy \& Performance}
\begin{figure}[t]
\begin{center}
\begin{tabular}{l}
\includegraphics [width=0.92\linewidth,height=8cm]{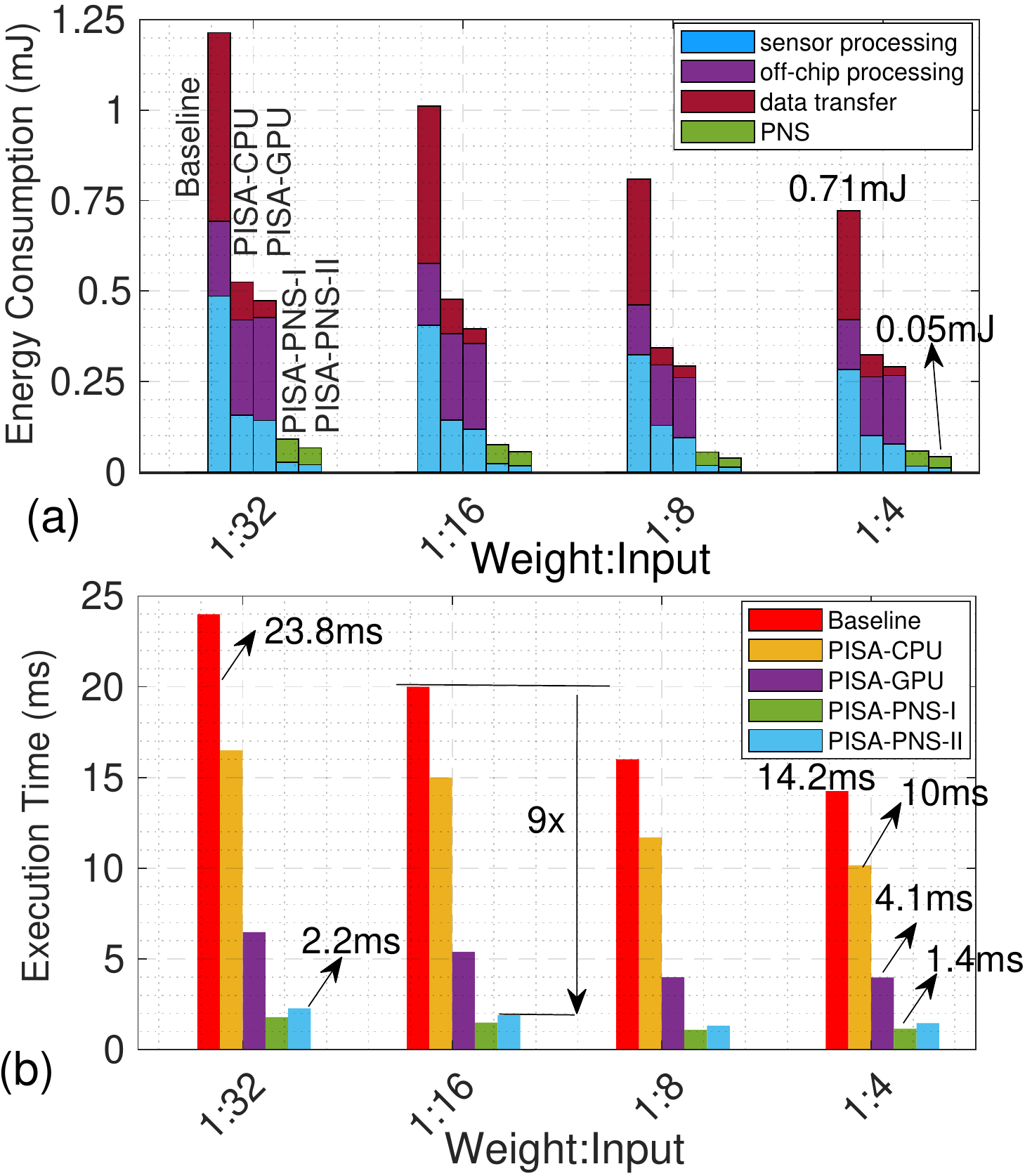} %\vspace{-0.9em}\\
 \end{tabular}
\caption{(a) Energy consumption, and (b) Execution time of under-test PISA-based platforms in various configurations compared with the baseline.}\vspace{-1em}
\label{perf}\vspace{-1em}
\end{center}
\end{figure}

We analyze the PISA's utility in processing the 1$^{st}$-Conv. layer for continuous mobile vision in three scenarios, i.e., assisting mobile CPU (PISA-CPU), assisting mobile GPU (PISA-GPU), and PISA-PNS, and compare it with a baseline sensor-CPU platform. For this goal, a BWNN model with 6 binary-weight Conv. layers and 2 FC layers to process the SVHN data-set is adopted. The energy consumption and latency results of the under-test platforms are then reported for four various weight/input configurations in PNS (W:I= 1:32, 1:16, 1:8, 1:4) in Fig. \ref{perf}. 
The under-test platforms in each experiment from left to right include the baseline design consisting of a conventional 128$\times$128 image sensor and an Intel(R) Core i7-6700 at 3.4GHz CPU with 16GB RAM where CPU plays the main role in processing all layers after receiving the raw data from the sensor's ADC.
The second platform consists of the same CPU connected to  128$\times$128 PISA array, where PISA processes 1$^{st}$ Conv. layer and remaining layers are processed by the CPU. The third design replaces the previous CPU with an NVIDIA GTX 1080Ti Pascal GPU with 3584 CUDA cores running at 1.5GHz (11TFLOPs peak performance). For CPU/GPU platforms, we use the open-source algorithm DoReFa-Net \cite{zhou2016dorefa} where the rest of the layers can be accelerated using the bit-wise convolution of fixed-point integers. 
The last two designs (fourth and fifth columns in each configuration in Fig. \ref{perf}) take advantage of PISA and its PNS-support to process the whole BWNN. When the 1$^{st}$ Conv. layer is processed by PISA, we adopted two alternative PIM techniques, i.e., DRISA \cite{li2017drisa} and our DRA mechanisms in the PNS unit to compute the 2$^{nd}$-6$^{th}$ Conv. and 2 FC layers near the sensor. Note that, any bit-wise PIM techniques could be adopted.

We report the breakdown of energy consumption into sensor processing, off-chip processing, data transfer, and PNS for various W:I configurations.
We find that PISA performs favorably against conventional CMOS image sensors. First, PISA substantially reduces the
data transmission energy by $\sim$84\%. paired with the CPU and GPU.
The PISA-CPU platform saves 58\% energy on average compared with the baseline as shown in Fig. \ref{perf}(a). While the PISA-GPU does not show a remarkable energy-saving over PISA-CPU but is still 89\% more energy-efficient than the baseline. Besides reduction in data transfer, the other reason behind such a striking energy saving is eliminating energy-hungry ADC units in PISA's processing mode.
Second, we observe that PISA-PNSs\footnote{PNS-I and PNS-II denote the DRISA-1T1C and our DRA mechanisms, respectively.} reduce the energy consumption of edge devices dramatically. The PISA-PNS-II requires $\sim$50-170$\mu$J energy depending on PNS configuration to process the whole BWNN on the edge, which is a safe choice for power-constrained IoT sensor devices. Please note that PISA-PNS designs almost eliminate the data transmission energy. Fig. \ref{perf}(b) illustrates the execution time corresponding to various W:I configurations. We observe that the PISA-PNS-II design achieves $\sim$3-7$\times$ speed-up in processing input frames compared with the baseline. However, PISA-PNS-I indicates a shorter execution time.

\subsection{Resource Utilization}
\begin{figure}[h]
\begin{center}
\begin{tabular}{l}
\includegraphics [width=0.98\linewidth]{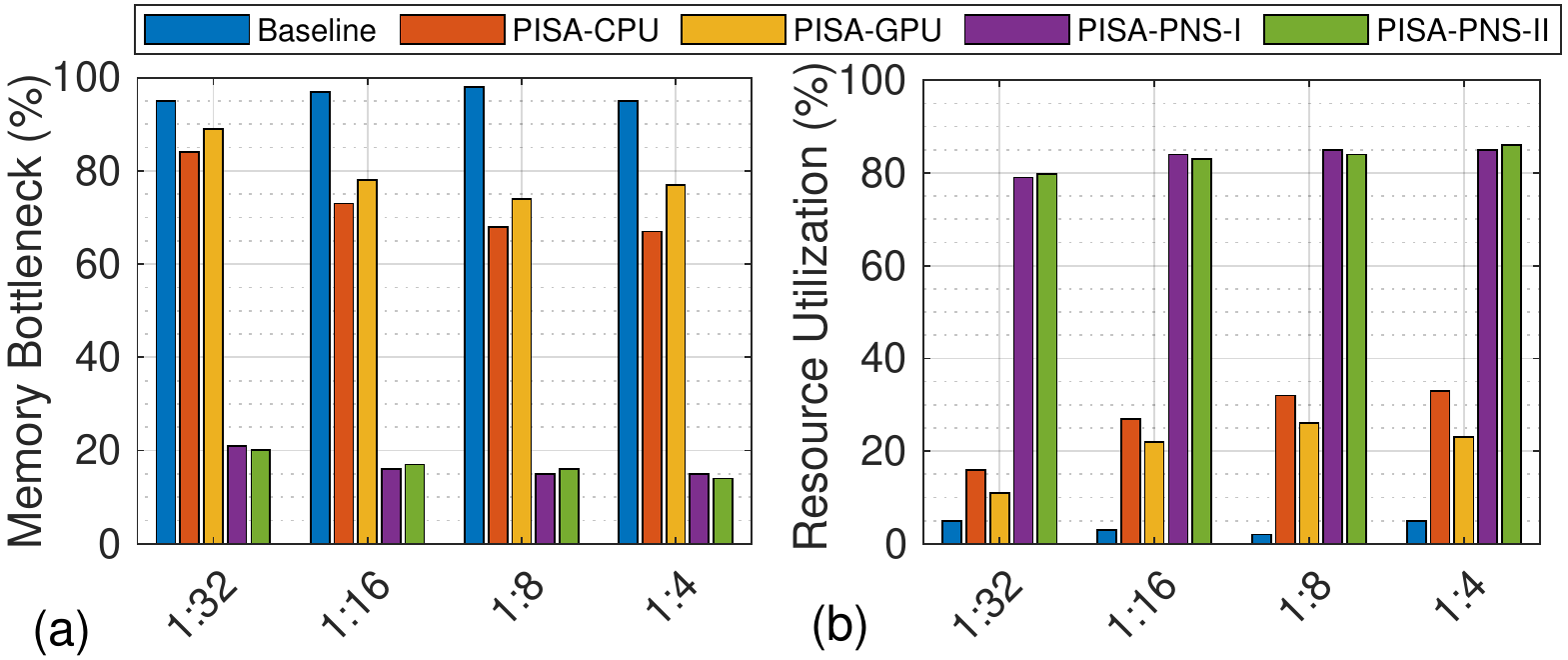} %\vspace{-0.9em}\\
 \end{tabular}
\caption{(a) Memory bottleneck ratio, (b) Resource Utilization ratio.}\vspace{-1em}
\label{MBR}\vspace{-1em}
\end{center}
\end{figure}

To explore the impact of PISA in reducing memory bottleneck in executing the $1^{st}$-layer of BWNN, we measured the time fraction at which on-/off-chip data transfer limits the performance as shown in Fig. \ref{MBR}(a). This evaluation was accomplished through experimentally extracted results of each platform with the number of memory access. We observe the PISA-PNS platforms spend less than 22\% of time for data conversion and memory access, whereas the baseline design spends over 90\% of its time waiting to load data from memory. A low memory bottleneck ratio can be translated to a high resource utilization ratio as depicted in Fig. \ref{MBR}(b). We observe that PISA-PNS platforms obtain the highest ratio utilizing up to 83\% computation resources. 

\begin{table*}[t]
\centering
\caption{Performance comparison of various PIS units.} %\vspace{-0.5em}
 \scalebox{0.85}{
\begin{tabular}{ccccccccccc}
\hline
 \textbf{Designs}              &  \textbf{\begin{tabular}[c]{@{}c@{}}Technology\\ ($nm$)\end{tabular}} &  \textbf{Purpose}                                                            &  \textbf{Comput. Scheme} &  \textbf{Memory} &  \textbf{NV*} &  \textbf{\begin{tabular}[c]{@{}c@{}}Pixel Size\\ ($\mu m^2$)\end{tabular}} &  \textbf{Array Size} &  \textbf{\begin{tabular}[c]{@{}c@{}}Frame Rate\\ ($frame/s$)\end{tabular}} &  \textbf{\begin{tabular}[c]{@{}c@{}}Power\\ ($mW$)\end{tabular}}               &  \textbf{\begin{tabular}[c]{@{}c@{}}Efficiency\\ ($TOp/s/W$)\end{tabular}} \\ \hline
\textbf{\cite{park20147}}     & 180                                                                & 2D optic flow est.                                                          & raw-wise                & Yes             & No           & 28.8$\times$28.8                                                          & 64$\times$64        & 30                                                                      & 0.029                                                                       & 0.0041                                                                  \\ \hline
\textbf{\cite{hsu20200}}      & 180                                                                & \begin{tabular}[c]{@{}c@{}}edge*/blur/sharpen/\\ 1st layer DNN\end{tabular} & raw-wise                & No              & No           & 7.6$\times$7.6                                                            & 128$\times$128      & 480                                                                     & \begin{tabular}[c]{@{}c@{}}sensing: 0.077 \\ processing: 0.091\end{tabular} & 0.777                                                                   \\ \hline
\textbf{\cite{yamazaki20174}} & 60/90                                                              & STP$^\dagger$                                                               & raw-wise                & Yes             & No           & 3.5$\times$3.5                                                            & 1296$\times$976     & 1000                                                                    & \begin{tabular}[c]{@{}c@{}}sensing: 230 \\ processing:363\end{tabular}      & 0.386                                                                   \\ \hline
\textbf{\cite{xu2020macsen}}  & 180                                                                & 1st layer BNN                                                               & entire-array            & No              & No           & 110$\times$110                                                            & 32$\times$32        & 1000                                                                    & 0.0121                                                                      & 1.32                                                                    \\ \hline
\textbf{\cite{carey2013100}}  & 180                                                                & edge*/TMF$^\ddagger$                                                        & raw-wise                & Yes             & No           & 32.6$\times$32.6                                                          & 256$\times$256      & 100,000                                                                 & 1230                                                                        & 0.535                                                                   \\ \hline
\textbf{PISA}                 & 65                                                                 & 1st layer BNN                                                               & entire-array            & Yes             & Yes          & 55$\times$55                                                              & 128$\times$128      & 1000                                                                    & \begin{tabular}[c]{@{}c@{}}sensing: 0.025\\ processing: 0.0088\end{tabular} & 1.745                                                                   \\ \hline
\end{tabular}}
\label{comp} %\vspace{-0.5em}
\end{table*}

\subsection{Comparison}
Table \ref{comp} compares the structural and performance parameters of selective PIS designs in the literature. As different designs are developed for specific domains, for an impartial comparison, we estimated and normalized the power consumption when all PIS units execute the similar task of processing the 1$^{st}$-layer of DNN. The PISA achieves the frame rate of 1000 and the efficiency of $\sim$1.745 TOp/s/W as the most efficient design. This comes from the massively-parallel CFP and eliminating ADC for coarse-grained detection. However, the design in \cite{carey2013100} achieves the highest frame-rate and the design in \cite{yamazaki20174} imposes the least pixel size enabling in-sensor computing. As for the area, our post-layout simulation results reported in Table I show a PISA’s compute-pixel occupies $\sim$55x55 $\mu m^2$ in 65nm. As we do not have access to the other layouts' configurations, it is very hard to have a fair comparison between area overheads. However, we believe a ballpark assessment can be made by comparing the number of minimum size transistors in previous SRAM-based designs and PISA’s lower-overhead compute add-on. We reimplemented MACSen \cite{xu2020macsen} at circuit-level as the only BWNN accelerator developed with the same purpose. Our evaluation showed that with the same near-sensor unit based on DRISA \cite{li2017drisa}, PISA consumes $\sim$40\% less power consumption. Putting everything together, PISA offers 
1) a low-overhead, dual-mode and reconfigurable design to keep the sensing performance and realize a processing mode to remarkably reduce the power consumption of data conversion and transmission;
2) single-cycle in-sensor processing mechanism to improve image processing speed;
3) highly parallel in-sensor processing design to achieve ultra-high-throughput; 
4) exploiting NVM which reduces standby power consumption during idle time and offers instant wake-up time, and resilience to power failure to achieve high performance.

\begin{figure}[t]
\begin{center}
%\begin{tabular}{l}
\includegraphics [width=\linewidth]{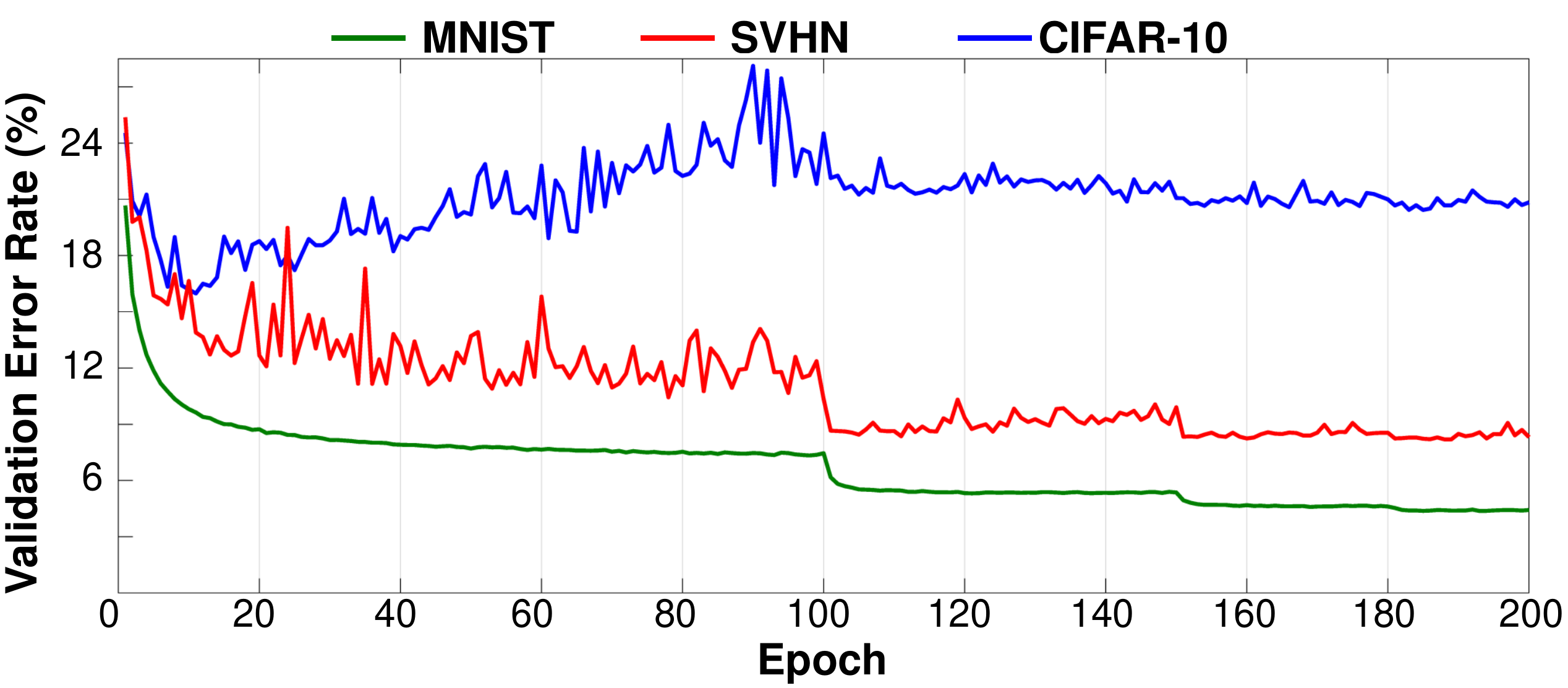} %\vspace{-0.75em}\\
% \end{tabular}%\vspace{-0.5em}
\caption{Validation error curves of three different datasets using the proposed BWNN configuration.}
\label{accuracyF}%\vspace{-1.75em}
\end{center}
\end{figure}

\subsection{Accuracy}
In the original BWNN topology, all the layers, except the first and last, are implemented with binarized weights \cite{BNN1,BNN2,FINN}. Since, in image classification tasks, the number of input channels is relatively smaller than the number of internal layers' channels, required parameters and computations are small. Thus, converting the input layer will not be a significant issue \cite{BNN1}. 
We conduct experiments on several datasets, including MNIST, SVHN, and CIFAR-10.
A BWNN model with 6 binary-weight Conv. layers and 2 FC layers to process the SVHN data-set is adopted. The $1^{st}$-layer consists of 32-by-32 images centered around a single character, where each pixel is mapped into a CP unit in PISA, the output states of PISA are then fed into the second layer implemented by near-memory design. 
Fig. \ref{accuracyF} shows the validation error versus the number of epochs of three different datasets in a worst-case scenario, i.e., with 1:4 configuration for 2$^{nd}$ to the last layer. The comparison of classification accuracy is summarized in Table \ref{accuracyT}. We find that the PISA shows an acceptable accuracy while providing significant energy-delay-product reduction as discussed earlier.

\begin{table}[h]
\centering
\caption{BNN accuracy (\%) on MNIST, SVHN and CIFAR-10.}
%\vspace{-0.7em}
\begin{tabular}{ccccc}
\hline
 \textbf{Configuration} &  \textbf{MNIST}  &   \textbf{SVHN}    &  \textbf{CIFAR-10}    \\ \hline 
\cite{BNN1}   &   96.0   &  97.47    &   89.85       \\ \hline
\cite{BNN2}   &   98.25    &  97.00     &   86.98        \\ \hline
\cite{FINN}    &   98.4    &  94.9     &   80.1        \\ \hline
\cite{FBNA}   &    --   &  96.9     &   88.61        \\ \hline
\textbf{Ours}   &  95.12   &   90.35  &  79.80    \\ \hline
\end{tabular}%\vspace{-0.5em}
\label{accuracyT}
\end{table}

\cmmnt{%%%%%%%%%%%%%%%%%%%%%%%%%%%%%%%%%%%%%%%%%
\subsection{Resilience to Power Failure}
Although almost all the state-of-the-art image sensor designs utilize effective methods to reduce dynamic energy consumption, including clock gating and low-voltage operation, an increasing number of modern intelligent sensors and more application scenarios, making the standby power dissipation of such systems a critical issue, which can limit the wider sensors’ applications. 
The emergence of energy harvesting systems as a promising approach for battery-less IoTs suffers from intermittent behavior, leading to data and environmental inconsistencies. For example, captured data by sensors become unstable if they are held for a long time without intermittent resilient architectures and/or harvestable sources. Moreover, since concurrency with sensors is relatively interrupt-driven, intermittency makes this concurrency control much more complex.
To solve the data consistency, PISA utilizes NVM elements, which reduces standby power consumption during idle time, instant wake-up time, and resilience to power failure, leading to high throughput and high performance at the cost of the minor accuracy degradation.
Due to the page limit, we plan to extend our future work to investigate image sensors' challenges in the presence of power failure for energy harvested systems, and more thoroughly discuss PISA's power failure resiliency in detail.}%%%%%%%%%%%%%%%%%%%%%%%%%%%%%%%%

\section{Discussion and Future Work}
Although almost all the state-of-the-art image sensor designs utilize effective methods to reduce dynamic energy consumption, including clock gating and low-voltage operation, an increasing number of modern intelligent sensors and more application scenarios, making the standby power dissipation of such systems a critical issue, which can limit the wider sensors’ applications. 
The emergence of energy harvesting systems as a promising approach for battery-less IoTs suffers from intermittent behavior, leading to data and environmental inconsistencies. For example, captured data by sensors become unstable if they are held for a long time without intermittent resilient architectures and/or harvestable sources. Moreover, since concurrency with sensors is relatively interrupt-driven, intermittency makes this concurrency control much more complex.
To solve the data consistency, PISA utilizes NVM elements, which reduces standby power consumption during idle time, instant wake-up time, and resilience to power failure, leading to high throughput and high performance at the cost of the minor accuracy degradation.
Due to the page limit, we plan to extend our future work to investigate image sensors' challenges in the presence of power failure for energy harvested systems, and more thoroughly discuss PISA's power failure resiliency.

% \textbf{Discussion and Future Work}
% Although almost all the state-of-the-art image sensor designs utilize effective methods to reduce dynamic energy consumption, including clock gating and low-voltage operation, an increasing number of modern intelligent sensors and more application scenarios, making the standby power dissipation of such systems a critical issue, which can limit the wider sensors’ applications. 
%Furthermore, the emergence of energy harvesting systems as a promising approach for battery-less IoTs suffers from intermittent behavior, leading to data and environmental inconsistencies. For example, captured data by sensors become unstable if they are held for a long time without intermittent resilient architectures or/and harvestable sources. Moreover, since concurrency with sensors is relatively interrupt-driven, intermittency makes this concurrency control much more complex. 
% To solve the data consistency, our proposed pixel utilized non-volatile memory elements, which reduce standby power consumption during idle time, instant wake-up time, and resilience to power failure, leading to high throughput and high performance at the cost of the minor accuracy degradation. Due to the page limit of this paper, we cannot thoroughly discuss PISA’s power failure resiliency in detail; thus, we plan to extend our future work to investigate image sensors’ challenges in the presence of power failure for energy harvested systems.

\section{Conclusion} 
In summary, this work proposed an efficient processing-in-sensor accelerator, namely PISA, for real-time edge-AI devices. PISA intrinsically performs a coarse-grained convolution operation on the 1$^{st}$-layer of binarized-weight neural networks leveraging a novel compute-pixel with non-volatile weight storage. The design was then completed by a near sensor processing-in-DRAM unit to perform a fine-grained convolution operation over the remaining layers.
Our results demonstrate acceptable accuracy on various data sets, where PISA achieves the frame rate of 1000 and the efficiency of $\sim$1.74 TOp/s/W.

\bibliographystyle{IEEEtran}%\vspace{-0.5em}

\end{document}